\documentclass[manuscript,screen]{acmart}
\AtBeginDocument{%
  }

\setcopyright{acmlicensed}
\copyrightyear{2026}
\acmYear{2026}
\acmDOI{XXXXXXX.XXXXXXX}

\acmVolume{TBD}
\acmNumber{TBD}
\acmArticle{TBD}

\acmISBN{978-1-4503-XXXX-X/2018/06}

\usepackage{algorithm}
\usepackage{algpseudocode}
\usepackage{subcaption}
\usepackage{amsmath}
\usepackage{booktabs}
\usepackage{multirow}
\usepackage{ulem}
\usepackage{comment}
\usepackage{graphicx}
\usepackage{amsfonts}
\usepackage{textcomp}
\usepackage{caption}
\usepackage{colortbl}
\usepackage{xcolor}
\usepackage{makecell}
\usepackage{appendix}
\usepackage{array} 

\newcommand{\figref}[1]{Figure~\ref{fig:#1}}   
\newcommand{\secref}[1]{Section~\ref{sec:#1}}
\newcommand{\tblref}[1]{Table~\ref{tab:#1}}

\newcommand{\algoref}[1]{Algorithm~\ref{alg:#1}}
\newcommand{\eqnref}[1]{Eq.~\ref{eq:#1}}

\begin{document}

\title{ARTA: Graph Combinatorial Optimization for Robust Interbeat Interval Estimation from Wearable Electrocardiogram Signals}

\author{Shovito Barua Soumma}
\orcid{0009-0007-1949-9795}
\email{shovito@asu.edu}
\affiliation{%
  \institution{Arizona State University}
  \city{Phoenix}
  \state{AZ}
  \country{USA}
}

\author{Asiful Arefeen}
\orcid{0000-0002-7876-3206}
\affiliation{%
  \institution{Arizona State University}
  \city{Phoenix}
  \state{AZ}
  \country{USA}
}

\author{Hassan Ghasemzadeh}
\orcid{0000-0002-1844-1416}
\affiliation{%
  \institution{Arizona State University}
  \city{Phoenix}
  \state{AZ}
  \country{USA}
}
\thanks{This work was supported in part by the National Science Foundation under grant 2531465. Any opinions, findings, conclusions, or recommendations expressed in this material are those of the authors and do not necessarily reflect the views of the funding organizations. Codebase: \href{https://github.com/shovito66/arta}{\textcolor{blue}{https://github.com/shovito66/arta}} and interactive dashboard: \href{https://arta.ghasemzadeh.com/}{\textcolor{blue}{https://arta.ghasemzadeh.com}}
}

\renewcommand{\shortauthors}{Soumma et al.}

\begin{abstract}
Accurate heart rate (HR) and interbeat interval (IBI) estimation from wearable electrocardiogram (ECG) signals is central to continuous cardiovascular and physiological monitoring in real-world environments. The research objective of HR/IBI estimation is to reliably identify ECG peaks that correspond to true cardiac beats, commonly referred to as R-peaks. However, this problem is highly challenging because motion artifacts introduced by user's physical activity create false beats and contaminate true R-peaks. 
\textcolor{black}{Current methods typically use a single peak-detection algorithm and make local corrections, such as removing unreasonably short intervals or filling excessively long ones. However, because these adjustments are made in isolation and lack a broader context of the heartbeat sequence, they often become unreliable when motion artifacts impact consecutive beats.}
To address this problem, we propose \textbf{ARTA} (Artifact-Robust Temporal Alignment), a combinatorial approach that models the problem of accurate HR/IBI estimation as a constrained sequence optimization task with the goal of selecting an ordered subset of candidate peaks that best explain the underlying heartbeat rhythm. ARTA models the ordering of the candidate peaks as a \textit{transition graph} and finds a minimum-cost path on the transition graph. The transition graph is constructed as a directed acyclic graph where graph nodes represent candidate peaks and graph edges represent valid transitions representing physiologically plausible RR intervals.
\textcolor{black}{Each candidate peak is scored based on detector agreement and whether it fits the recent IBI rhythm. Beam-search dynamic programming then selects the path that best satisfies these criteria while spanning the full recording rather than only a clean segment. As a result, artifact-induced false peaks are rejected when they disrupt the rhythm, and missed beats can be recovered when another detector proposes a plausible peak nearby.}
On ECG recordings corrupted with realistic electrode-motion artifact across six SNR levels ($-2$ to 12~dB), ARTA achieves a mean absolute error of 1.796 BPM (54\% improvement over the best single-detector baseline) and places 92.38\% of 8-second estimation windows within ±5 BPM, with an R-peak F1 score of 0.95 and an IBI estimation error of 6.52 ms, substantially outperforming single-detector, naive-fusion, deep-learning, and prior graph-based baselines — with the largest gains under the most severe noise. Being training-free, interpretable, and single-channel, ARTA offers a practical foundation for trustworthy HRV-grade heartbeat tracking on resource-constrained wearable devices, running $\approx27\times$ faster than real time with a peak memory footprint below 2.1~MB in a single-thread CPU edge simulation.

\end{abstract}


  
\maketitle 

\section{Introduction}


In the last decade, cardiovascular diseases (CVDs) have become the leading cause of death globally, responsible for approximately 17.9 million deaths annually~\cite{WHO2017,mcaloon2016changing}. Continuous cardiac monitoring is therefore important for early risk assessment, disease management, and long-term evaluation of cardiovascular function, as variations in heart rhythms can provide early indications of heart attack, stroke, and other medical complications. Beyond clinical practice, the measurement of interbeat intervals (IBIs) and heart rate variability (HRV) is vital for assessing cognitive load in specialized fields like surgery and driving~\cite{meteier2021classification,meteier2022relevant}, as well as monitoring stress and sleep apnea~\cite{zhou2020multimodal}. While conventional monitoring often requires labor-intensive hospital visits, wearable sensors offer a promising alternative for continuous, non-invasive, and objective assessment in real-time daily life~\cite{Aygun}. 

Among the cardiac measures obtained from wearable signals, heart rate (HR), interbeat interval (IBI), and heart-rate variability (HRV) are especially important because they reflect both average cardiac activity and beat-to-beat autonomic regulation. Low HRV is specifically associated with higher risk of all-cause death and cardiovascular events in patients with established CVD~\cite{fang2020heart, huang2021graph}.  Prior work has emphasized that IBI and HRV can provide useful information for cardiovascular monitoring, stress assessment, emotion recognition, sleep quality, and other health-related applications~\cite{correia2023associations, coutts2020deep, bulagang2021novel}. Accurate IBI estimation is therefore the foundational computational step for a broad and growing range of clinical and consumer health applications.

Accurate IBI estimation requires reliable localization of consecutive heartbeats. In ECG, this is typically achieved by detecting R-peaks and measuring the time interval between adjacent peaks. However, ECG recordings collected during physical activity are frequently corrupted by motion artifacts, muscle activity, electrode displacement, baseline wander, and other noise sources. These distortions can substantially alter the morphology of the ECG waveform and make R-peak detection unreliable. In particular, electrode-motion artifact is difficult to remove because its frequency content may overlap with the PQRST complex, limiting the effectiveness of simple filtering approaches~\cite{tiramisu, malghan2020review}. 
Both ECG and PPG signals are acutely vulnerable to artifact induced by body movement during physical activity.  For PPG, the pulsatile component captured by wrist-worn sensors has inherently small amplitude, and even moderate physical activity induces sliding and rubbing effects on the optical sensor that severely degrade morphological signal quality~\cite{Aygun}. Importantly, estimating HRV under these conditions is fundamentally harder than estimating average heart rate. Average HR changes slowly and consistently, making it separable from noise through spectral or averaging methods. HRV and motion artifact are both temporally inconsistent, occupying the same statistical regime and rendering them extremely difficult to disentangle~\cite{Aygun,huang2021graph}.

\begin{figure}[h]
    \centering
    \includegraphics[width=1\linewidth, trim={5 20 4 55},clip]{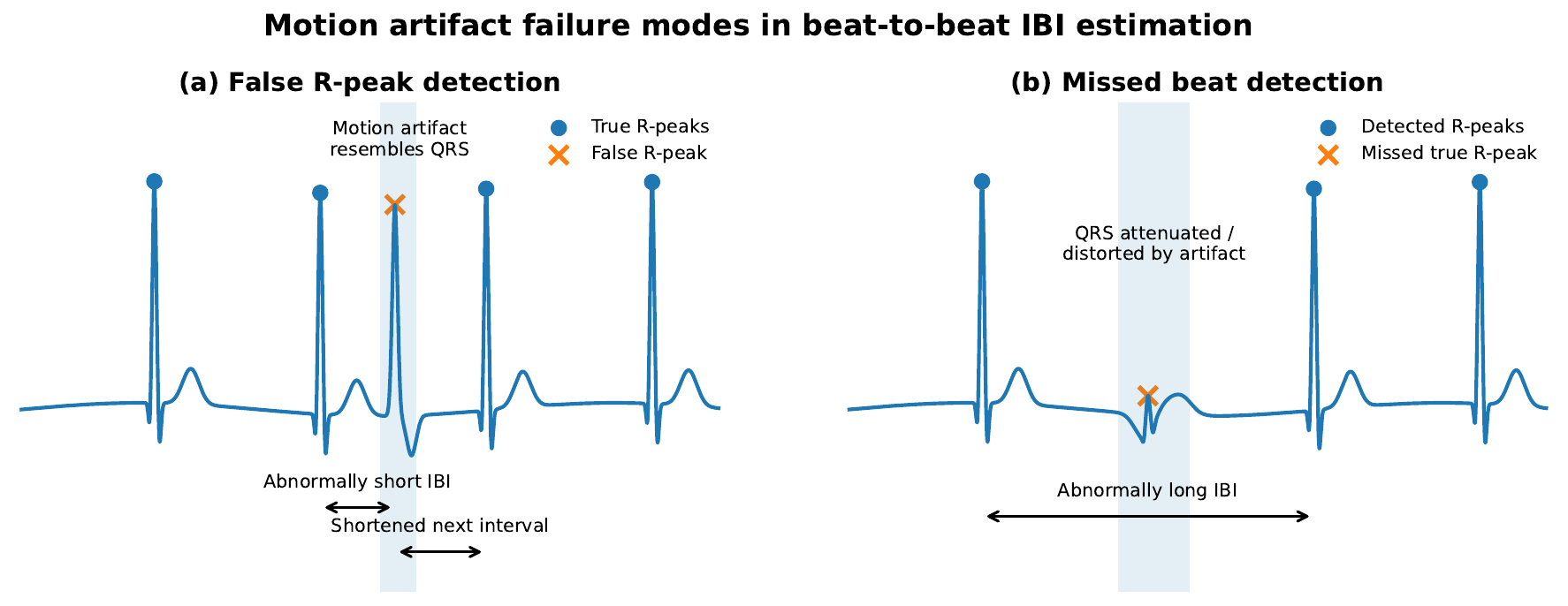}
    \caption{Motion-Induced Failure Modes in IBI Estimation}
    \label{fig:motion-effect}
\end{figure}
\begin{figure*}[t]
    \centering
    \includegraphics[width=0.85\linewidth, trim={0 0 0 30},clip]{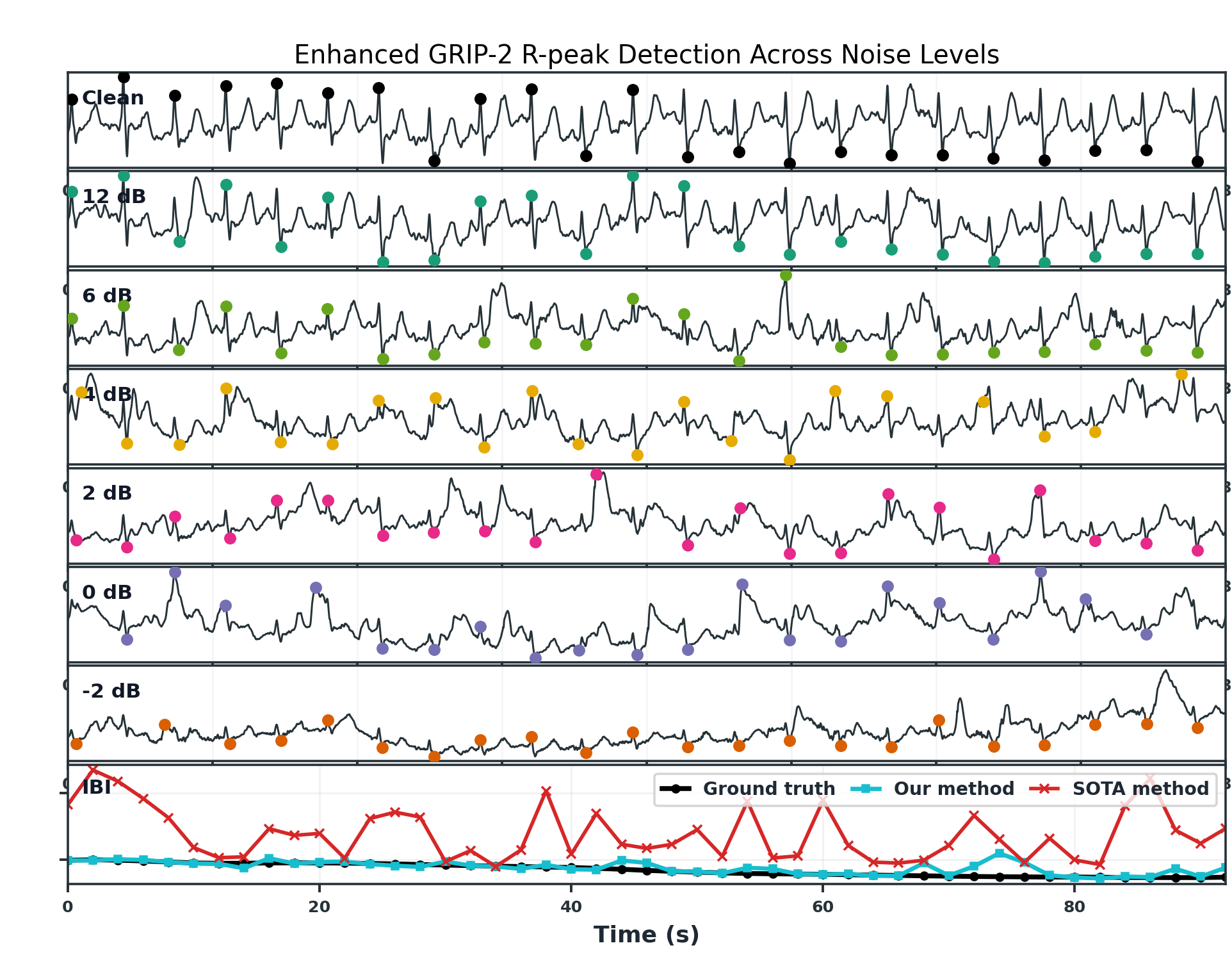}
    \caption{A clean ECG segment and its noisy variants. EM, MA, and BW noise have been added to the clean signal; the labels at the left of each row indicate the corresponding SNR. Colored markers show R-peaks detected at each noise level. The bottom row compares IBI sequences over time: IBIs estimated from noisy ECG by a state-of-the-art method (\textcolor{red}{red}) deviate substantially from the reference (black), whereas the proposed method (\textcolor{cyan}{cyan}) remains close to the ground truth across the recording.}
    \label{fig:ecg_vs_noise}
\end{figure*}

Motion artifact affects IBI estimation through two coupled failure modes (shown~\figref{motion-effect}). First, artifact-induced deflections may resemble QRS complexes and be falsely detected as R-peaks, producing abnormally short RR intervals. Second, true QRS complexes may be attenuated or distorted, producing missed beats and abnormally long RR intervals. These errors directly propagate into IBI, HRV, and window-level HR estimates. 
The severity of this degradation is illustrated in~\figref{ecg_vs_noise}, which shows a representative clean ECG segment alongside its noisy variants at progressively lower SNR levels: as motion artifact intensifies, conventional ECG detectors localize R-peaks inconsistently from one noise level to the next, and the IBI sequence produced by a state-of-the-art method fluctuates sharply and drifts away from the true IBI trajectory, whereas the proposed method tracks the reference closely throughout.
This makes robust IBI estimation more challenging than average HR estimation: average HR can often be estimated over longer windows, whereas IBI and HRV depend on precise beat-to-beat timing. Prior wearable-sensing work has similarly noted that IBI and HRV estimation deteriorate substantially under motion artifact and are more difficult than average HR estimation during intensive activity.


Existing methods for IBI estimation under motion artifact each address the problem only partially. Conventional R-peak detectors evaluate candidate peaks independently with no model of beat-to-beat regularity, making them brittle when artifact resembles cardiac morphology. Rule-based post-hoc repair corrects anomalous intervals locally but lacks global sequence context, leaving correlated artifact patterns undetected~\cite{10816301}. Deep learning approaches can suppress severe noise but require large paired training datasets, generalize poorly to unseen artifact distributions, and carry computational costs unsuitable for wearable deployment~\cite{tiramisu}. Graph-based methods offer a more principled alternative by encoding heartbeat continuity as a shortest-path problem~\cite{Aygun,huang2021graph}, but prior formulations tie edge weights to a global average HR estimate that degrades under the low-SNR conditions where robust IBI estimation is most needed — biasing the optimizer toward incorrect paths~\cite{Aygun}. Simple detector fusion or majority voting does not resolve this gap, because evaluating candidates independently ignores whether the selected sequence forms a physiologically plausible rhythm: a candidate detected by only one algorithm may still be a true missed beat, while a morphologically strong candidate may still be false if it creates an implausible RR interval. These limitations motivate a structured ECG-only framework that treats multiple detector outputs as complementary weak evidence and selects the final R-peak sequence through global, rhythm-aware path optimization.

\begin{figure*}[t]
    \centering
    \includegraphics[width=1\linewidth, trim={5 18 3 18},clip]{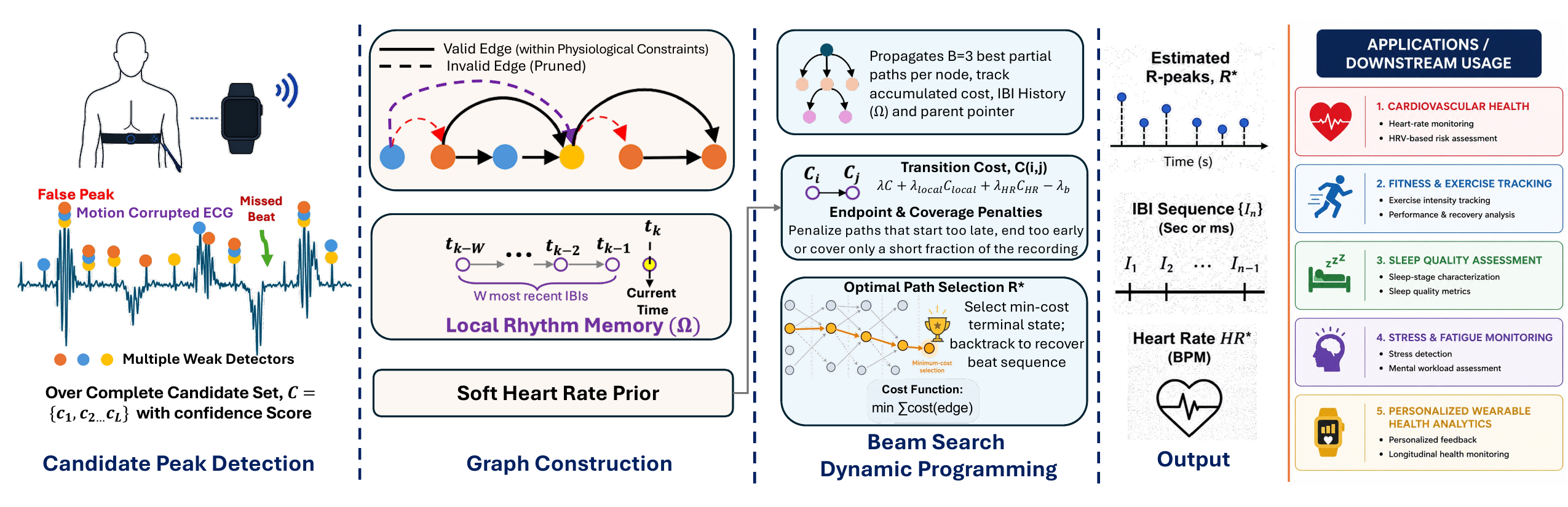}
    \caption{Overview of ARTA Framework. Multiple ECG detectors generate an over-complete set of candidate R-peaks, which are organized into a directed acyclic graph (DAG). Local rhythm memory, a time-varying heart-rate prior, and candidate confidence scores define transition costs. Beam-search dynamic programming identifies the minimum-cost heartbeat sequence, producing robust R-peak, IBI, and heart-rate estimates for downstream physiological monitoring applications.}
    \label{fig:framework}
\end{figure*}
To address these limitations, we propose \textbf{A}rtifact-\textbf{R}obust \textbf{T}emporal
\textbf{A}lignment (ARTA)\footnote{Interactive dashboard:
\href{https://arta.ghasemzadeh.com/}{\textcolor{blue}{https://arta.ghasemzadeh.com}},  Codebase: \href{https://github.com/shovito66/gcsd}{\textcolor{blue}{https://github.com/shovito66/arta}}}, an ECG-only framework for robust R-peak, IBI, and HR estimation under motion artifact. ARTA treats the outputs of multiple ECG detectors as weak and noisy observations of an unknown physiological heartbeat sequence. Candidate R-peaks are generated from WFDB-XQRS, Pan-Tompkins, and wavelet-based detection, merged into an over-complete candidate set, and represented as nodes in a directed acyclic graph. Edges encode physiologically feasible RR transitions. A beam-search dynamic programming algorithm then selects the globally optimal heartbeat path using a composite objective that combines local rhythm consistency, soft agreement with a time-varying HR prior, candidate confidence, beat reward, and endpoint/coverage penalties.

The key idea is that false-positive rejection and missed-beat recovery should be solved jointly as a sequence optimization problem rather than as independent peak-level decisions. In ARTA, false candidates are rejected when their inclusion increases rhythm inconsistency or candidate cost, while missed beats can be recovered when auxiliary detector candidates reduce long-gap penalties and improve global path consistency. Thus, the final R-peak sequence is selected not only because individual peaks appear reliable, but because the entire path forms a physiologically coherent heartbeat sequence.

The main contributions of this work are as follows:
\begin{enumerate}
    \item A graph combinatorial optimization formulation is introduced for robust ECG IBI estimation under motion artifact, where outputs from multiple R-peak detectors are treated as weak evidence rather than final decisions.
    \item An over-complete candidate directed acyclic \textit{transition graph} is constructed from complementary ECG detectors, with candidate relationships modeled through physiologically feasible RR transitions.
    \item A local-rhythm-aware dynamic programming objective is introduced that leverages path-specific IBI history rather than relying on average-HR guidance alone.
    \item Detector confidence, beat reward, and endpoint/coverage penalties are integrated into the path objective to suppress false positives, recover detector-missed beats when auxiliary candidates are available, and avoid short-fragment solutions.
    \item The proposed ARTA framework is evaluated across multiple motion-artifact SNR levels using both window-level HR metrics and beat-level diagnostic metrics.

\end{enumerate}

 
The remainder of this paper is organized as follows. Section~II reviews related work on QRS detection and graph-based IBI estimation. Section~III describes the ARTA framework in full mathematical detail. Section~IV presents experimental methodology. Section~V reports quantitative results at multiple noise levels, including beat-level, window-level, and HRV-parameter evaluation, an ablation study, and an edge-device feasibility analysis on a single-thread CPU simulation. Section~VI discusses the implications of the results and limitations of the method. Section~VII concludes.
\section{Related Work}
\label{sec:relatedworks}
Cardiac monitoring relies on several signal modalities, each with distinct trade-offs between signal fidelity and practical deployability. 
Various methods have been developed for heart rate (HR), interbeat interval (IBI), and heart-rate variability (HRV) estimation from ECG, PPG, and unobtrusive sensing modalities. ECG remains the standard modality for beat-level cardiac analysis because R-peaks provide a direct fiducial marker for estimating RR intervals~\cite{gao2025signal}. PPG, in contrast, is more convenient for wearable monitoring but is highly vulnerable to motion artifact, sensor displacement, skin-contact variation, and morphology distortion, making IBI and HRV estimation more difficult than average HR estimation~\cite{gao2025signal,bai2024hsf}. Beyond ECG and PPG, unobtrusive modalities such as ballistocardiography (BCG), Doppler cardiography (DCG), and other cardiac vibration signals have also been explored for long-term monitoring; however, these signals often lack clear ECG-like R-wave morphology and vary significantly across subjects, sensors, and postures~\cite{inan2009robust,jung2021accurate}.

\subsection{Traditional Detection Methods \& Signal Processing}
Traditional ECG-based R-peak detectors rely on filtering, differentiation, adaptive thresholding, and time-frequency analysis. Representative methods include the Pan-Tompkins detector~\cite{pan1985real}, WFDB-based QRS detectors such as XQRS/GQRS~\cite{silva2014open,10176714, editionwfdb}, and wavelet-based ECG delineation or QRS detection~\cite{martinez2004wavelet}. These algorithms are computationally efficient and interpretable, which makes them attractive for wearable implementation. However, they typically make local peak-level decisions and can fail when motion artifact creates QRS-like false peaks or distorts true QRS complexes. In the proposed framework, these methods are therefore used as weak detectors rather than final decision makers: WFDB-XQRS, Pan-Tompkins, and wavelet-based detection generate an over-complete candidate set, while the final heartbeat sequence is selected by global graph optimization.

A large body of work has also focused on PPG-based HR estimation under motion artifact. Signal-processing methods such as TROIKA~\cite{zhang2014troika}, WFPV~\cite{temko2017accurate}, particle filtering~\cite{nathan2015particle}, adaptive filtering, empirical mode decomposition, and spectral peak tracking have shown strong performance for average HR estimation during exercise. More recent approaches incorporate motion and historical HR information. Xiong et al. formulated spectral peak selection as an SVM classification problem using features such as peak coefficient ratio and distance from the previous HR peak~\cite{xiong2017svm}. Meng et al. used acceleration-derived step-frequency information and historical HR to constrain the candidate HR search space~\cite{ronghao2022heart}. Lan et al. combined triaxial-acceleration-based motion assessment, RLS filtering, empirical wavelet transform, and finite-state-machine reliability evaluation for PPG HR estimation~\cite{lan2024adaptive}. Guo et al. further integrated SVM-based signal-quality assessment and spectral peak selection for dynamic PPG HR extraction~\cite{guo2026research}. 

While effective for window-level HR, these frequency-domain approaches cannot resolve beat-to-beat IBIs, which require substantially finer temporal precision. Post-hoc heuristic repair, which interpolates over long gaps and discards short intervals, partially compensates for single-detector errors but remains local and fails when the artifact spans multiple consecutive beats.

~\subsection{Deep Learning and Hybrid Physiological Models}
Machine-learning and deep-learning methods have also been explored for ECG/PPG denoising, HR prediction, and downstream physiological inference. Arefeen et al. proposed a modified Tiramisu autoencoder to suppress ECG motion artifact and make R-peaks more prominent before IBI estimation~\cite{tiramisu}. Complementary architectures include bidirectional LSTMs for noise suppression~\cite{laitala2020robust}, CNN encoder-decoders for QRS complex recovery in armband-worn ECG~\cite{reljin2020using}, U-Net-based regression models that reformulate peak detection as a distance-transformation task~\cite{vijayarangan2020rpnet}, and multi-stage pipelines combining wavelet CNN autoencoders with ConvLSTM peak classifiers~\cite{yuen2020detecting}. Generative adversarial networks have also been applied to synthesize training data for ECG denoising~\cite{antczak2020generative}.

Nazaret et al. combined physiological ordinary differential equation modeling with neural networks to learn personalized HR responses from large-scale wearable exercise data~\cite{nazaret2023modeling}. This work exemplifies the power of embedding physiological structure into large-scale wearable data modeling. HR and IBI features have also been used in affective computing; for example, Bulagang et al. used SVM classification with combined HR and IBI features for emotion recognition in virtual reality~\cite{bulagang2021novel}. Although AI-based methods can capture complex nonlinear patterns, they often require labeled training data, may generalize poorly to unseen artifact distributions, and do not always provide interpretable beat-sequence correction.

\subsection{Graph-Based and Probabilistic Methods}
Graph-based IBI estimation is most closely related to the proposed method. Aygun et al. represented candidate morphological features as graph vertices and candidate IBIs as edges, then used shortest-path optimization to estimate IBI and HRV from motion-corrupted wearable signals~\cite{Aygun}. Huang et al. extended this idea to multi-channel PPG during intensive activity by constructing a directed acyclic graph from candidate heartbeat features and solving IBI estimation as a shortest-path problem~\cite{huang2021graph}. These methods demonstrate that heartbeat continuity can be effectively modeled through graph optimization. However, prior graph methods rely strongly on average-HR or average-IBI guidance for edge weighting. When this prior becomes unreliable under severe motion artifact, the graph may be biased toward incorrect paths.

The proposed ARTA framework builds on prior work while addressing its core limitations. It moves beyond single-detector pipelines by evaluating whether selected candidates collectively form a physiologically plausible rhythm, not merely whether each peak looks like a QRS in isolation. It improves on simple detector fusion by optimizing over the full beat sequence rather than scoring candidates independently. And it advances prior graph-based methods by replacing global average-HR edge weights with path-specific local IBI history maintained through beam-search dynamic programming, eliminating dependence on an external HR prior that degrades precisely where robustness is most needed. The result is an ECG-only framework that jointly resolves false-positive rejection and missed-beat recovery as a single global sequence optimization problem.

\section{Method}
\label{sec:method}
Our proposed method can be viewed as a graph-based heartbeat tracking framework illustrated in~\figref{framework}. Multiple ECG detectors first generate an over-complete set of candidate R-peaks containing both true beats and artifact-induced false peaks. These candidates are represented as nodes in a directed acyclic graph (DAG), where edges connect physiologically plausible consecutive beats satisfying RR interval constraints.

Rather than detecting peaks independently, the method searches for the optimal path through the graph. The path cost is designed to favor locally consistent rhythm, agreement with the heart-rate trend, high-confidence candidates, and paths that explain the full ECG recording. As a result, false-positive rejection and missed-beat recovery emerge naturally from global path optimization.

\subsection{Problem Statement}
\label{sec:problem}

Let $x[n]$, $n = 1,\ldots,N$, denote a noisy ECG signal sampled
at frequency $f_s$, and let $\mathcal{R} = \{r_1, r_2, \ldots, r_M\}$
denote the unknown true R-peak sequence. Motion artifact corrupts
$x[n]$ such that $\mathcal{R}$ cannot be recovered directly from
standard detection. The goal is to find an estimated R-peak sequence
\begin{equation}
    \hat{\mathcal{R}} = \{\hat{r}_1, \hat{r}_2, \ldots, \hat{r}_M\}
\end{equation}
that minimizes the deviation from the true sequence in terms of
interbeat interval error:
\begin{equation}
    \hat{\mathcal{R}}^{*} = \arg\min_{\hat{\mathcal{R}}\,\in\,\mathcal{F}}
    \sum_{k=1}^{M-1} \left|\hat{I}_k - I_k\right|,
\end{equation}
where $\hat{I}_k = t(\hat{r}_{k+1}) - t(\hat{r}_k)$ is the estimated
IBI, $I_k = t(r_{k+1}) - t(r_k)$ is the true IBI, and $\mathcal{F}$
denotes the feasible set of physiologically plausible beat sequences
satisfying
\begin{equation}
    \tau_{\min} \leq \hat{I}_k \leq \tau_{\max}
    \quad \forall\, k,
\end{equation}
where $\tau_{\min}$ and $\tau_{\max}$ are the minimum and maximum
admissible RR intervals, respectively. Since the true sequence
$\mathcal{R}$ is unobservable from the noisy signal, the true IBI
error cannot be minimized directly. ARTA approximates this objective
by constructing a surrogate cost $J(\hat{\mathcal{R}})$ that penalizes
rhythm inconsistency, deviation from a heart-rate prior, and
low-confidence candidates -- all computable from $x[n]$ alone:
\begin{equation}
    \hat{\mathcal{R}}^{*} = \arg\min_{\hat{\mathcal{R}}\,\in\,\mathcal{F}}
    \;J(\hat{\mathcal{R}}).
\end{equation}
The surrogate cost $J(\hat{\mathcal{R}})$ is defined in full in~\secref{transition_cost}. The estimated IBI sequence and
window-level heart rate follow directly as
\begin{equation}
    \hat{I}_k = t(\hat{r}_{k+1}) - t(\hat{r}_k),
    \qquad
    \widehat{\mathrm{HR}}_k = \frac{\gamma}{\hat{I}_k},
\end{equation}
where $\gamma$ is the millisecond-to-BPM conversion constant.
Evaluation is performed in windows of duration $T_w$ with stride $T_s$ to align with the reference BPM measurements.

\subsection{Candidate R-Peak Generation}
The first stage of ARTA converts the noisy ECG signal $x[n]$ into an over-complete set of candidate R-peaks. Rather than relying on a single detector, ARTA uses multiple ECG-only detectors as weak estimators of the latent heartbeat sequence. This design increases candidate coverage: a true beat missed by one detector may still be proposed by another, while false detections can later be rejected by the graph optimization stage.
WFDB-XQRS provides the strongest single-detector backbone,
$C_{\mathrm{XQRS}}$, owing to its robust wavelet-based QRS localization;
Pan--Tompkins contributes $C_{\mathrm{PT}}$, offering complementary
sensitivity to QRS energy through its differentiation-and-integration
pipeline; and a wavelet-domain detector contributes $C_{\mathrm{WT}}$,
improving morphological selectivity in distorted segments.
The outputs of these detectors are not treated as final R-peak decisions; instead, they form a noisy evidence pool from which ARTA selects the most physiologically consistent heartbeat sequence. The three sets are then merged using non-maximum suppression (NMS) with a minimum separation equal to a physiological refractory period, yielding:

\begin{equation}
    \mathcal{C} = \mathrm{NMS}\!\left(C_{\mathrm{XQRS}} \cup C_{\mathrm{PT}} \cup C_{\mathrm{WT}}\right)
\end{equation}
After merging, the candidates are sorted by time:
\small{\(t(c_{1}) < t(c_{2}) < \cdots < t(c_{L})\)}. The merged set is intentionally over-complete, containing both true R-peaks  and artifact-induced false candidates \small {($\mathcal{C} = R_{true} \cup A_{false}$)}. The role of the downstream optimization is to separate these two components in order to select a physiologically plausible subsequence from this over-complete set.

\subsection{Candidate Confidence Scoring}
Each candidate \(c_j\) is assigned a scalar confidence score \(q_j \in [0,1]\), which reflects both detector agreement and local ECG evidence. The confidence score is designed to favor candidates supported by the strongest detectors while still allowing auxiliary candidates to participate in the graph optimization when they improve rhythm consistency. The confidence score is defined as
\begin{equation}
q_j = \alpha_x d_x(j) + \alpha_p d_p(j) + \alpha_w d_w(j) + \alpha_a A_j,
\end{equation}
where \small{\(d_x(j), d_p(j), d_w(j) \in \{0,1\}\)} indicate agreement with the XQRS, Pan--Tompkins, and wavelet detectors, respectively, and \(A_j \in [0,1]\) is a normalized local ECG quality score. The weights satisfy
\begin{equation}
\alpha_x, \alpha_p, \alpha_w, \alpha_a \geq 0,
\qquad
\alpha_x + \alpha_p + \alpha_w + \alpha_a = 1.
\end{equation}

Because XQRS provides the strongest ECG-only backbone in the proposed framework, its weight is typically chosen such that
\begin{equation}
\alpha_x > \alpha_p, \alpha_w .
\end{equation}
After scoring, the confidence is clipped to the interval \([0,1]\):
\begin{equation}
q_j \leftarrow \operatorname{clip}(q_j,0,1).
\end{equation}
The node penalty used in graph optimization is then defined as
\begin{equation}
C_{\mathrm{node}}(j) = 1 - q_j .
\end{equation}
Thus, high-confidence candidates incur low node cost, while lower-confidence candidates remain available to the optimizer if they improve local rhythm consistency or global path continuity.

\subsection{Directed Acyclic Graph (DAG) Construction}

A directed acyclic graph is constructed as $G=(V,E)$ where \small{$V=\mathcal{C}$}. A directed edge from candidate \(c_i\) to candidate \(c_j\), where \(i < j\), is allowed only if the resulting RR interval is physiologically plausible:
\begin{equation}
\tau_{\min} \leq w_{ij} \leq \tau_{\max},
\end{equation}
where $w_{ij} = t(c_j) - t(c_i)$, and $\tau_{\min}$, $\tau_{\max}$ 
are the minimum and maximum admissible RR intervals defined in~\secref{problem}.
As all edges move forward in time, the resulting structure is a DAG, enabling efficient dynamic programming with guaranteed termination.

\begin{algorithm}[t]
\caption{Candidate graph construction}
\label{alg:gcsd-graph}
\begin{algorithmic}[1]
\Require ECG $x[n]$, $f_s$, $\mathrm{\tau}_{\min}$, $\mathrm{\tau}_{\max}$,
         detector weights $w_x, w_p, w_w, w_a$
\Ensure Candidate set $\mathcal{C}$, confidence $\{q_j\}$,
        predecessor lists $\{\mathrm{Pred}[j]\}$
\For{each detector $d \in \{\text{XQRS, PT, WT}\}$}
    \State $\mathcal{C}_d \leftarrow \mathrm{Apply}_d(x[n])$
\EndFor
\State $\mathcal{C} \leftarrow \mathrm{NMS}(\mathcal{C}_\text{XQRS}
       \cup \mathcal{C}_\text{PT} \cup \mathcal{C}_\text{WT})$;
       \;\text{sort } $\mathcal{C}$ \text{ by time}
\For{each candidate $c_j \in \mathcal{C}$}
    \State $q_j \leftarrow w_x\mathbf{1}_{[j\in\text{XQRS}]}
            + w_p\mathbf{1}_{[j\in\text{PT}]}
            + w_w\mathbf{1}_{[j\in\text{WT}]}
            + w_a A_j$
    \State $C_\text{node}(j) \leftarrow 1-q_j$
    \State $\mathrm{Pred}[j] \leftarrow
           \{(i,\,w_{ij}) : i<j,\;
           \mathrm{\tau}_{\min}\le w_{ij}\le \mathrm{\tau}_{\max}\}$
\EndFor
\State \Return $\mathcal{C},\;\{q_j\},\;\{\mathrm{Pred}[j]\}$
\end{algorithmic}
\end{algorithm}

\subsection{Local Rhythm Memory}
\label{sec:local_rhythm}
For each partial path ending at candidate \(c_i\), the algorithm maintains a recent IBI window
\[
\Omega_i = \{I_{i-W+1}, \ldots, I_i\},
\]
where \(W\) denotes the local rhythm memory length. The local rhythm reference is defined as
\[
m_i = \operatorname{median}(\Omega_i),
\]
and the local robust variability estimate is
\[
s_i = \operatorname{MAD}(\Omega_i)
= \operatorname{median}_{I \in \Omega_i}
\left( \lvert I - m_i \rvert \right).
\]

For a transition \(i \rightarrow j\), the local rhythm penalty is
\begin{equation}
C_{\mathrm{local}}(i,j)
=
\begin{cases}
0,
& |\Omega_i| < W_{\min},
\\[6pt]
\dfrac{
\left| w_{ij} - m_i \right|
}{
s_i + \epsilon
},
& |\Omega_i| \geq W_{\min}.
\end{cases}
\end{equation}
where \(\epsilon\) is a small positive constant preventing division by zero. When path history is too short (\small{\(|\Omega_i|<W_{min}\)}) to form a reliable estimate during the early stage of a path, \small{\(C_{\mathrm{local}}(i,j) = 0\)}. Using the path-specific window \(\Omega_i\) rather than a global average allows the rhythm reference to adapt to the actual pace of each candidate path, thereby correctly handling both resting heart rate and exercise-induced tachycardia without parameter re-tuning. This term is the core physiological constraint of the method. It measures how well a new RR interval agrees with the recent rhythm trajectory of the current path.

\subsection{Time-Varying Heart-Rate Prior}
A soft time-varying HR prior, \(\hat{h}(t)\), is estimated from preliminary ECG-only detections in overlapping windows and converted to a prior IBI as
\[
\hat{I}(t) = \frac{\gamma}{\hat{h}(t)}
\]
 For an edge ending at candidate \(c_j\), the HR-prior penalty is
\[
C_{\mathrm{HR}}(i,j)
=
\frac{
|w_{ij} - \hat{I}(t_j)|
}{
\sigma_{\mathrm{HR}}
},
\]
where \(t_j = t(c_j)\), and \(\sigma_{\mathrm{HR}}\) controls tolerance to prior mismatch.

The prior is deliberately soft. It guides the graph when local rhythm is uncertain, but it does not force the final path to follow the coarse HR estimate. If the prior estimate is poor at low SNR, the local rhythm cost $C_{local}$ continues to provide reliable guidance from the path's own history, rather than propagating prior estimation errors into the final result.

\subsection{Transition Cost}
\label{sec:transition_cost}
For a transition from \(c_i\) to \(c_j\), the total transition cost is
\begin{equation}
\small
C(i,j)
=
\lambda_{\mathrm{local}} C_{\mathrm{local}}(i,j)
+
\lambda_{\mathrm{HR}} C_{\mathrm{HR}}(i,j)
+
\lambda_{\mathrm{node}} C_{\mathrm{node}}(j)
-
\lambda_b
\end{equation} 

where \(\lambda_{\mathrm{local}}\) controls local rhythm consistency, \(\lambda_{\mathrm{HR}}\) controls HR-prior influence, \(\lambda_{\mathrm{node}}\) controls candidate-confidence influence, and \(\lambda_b>0\) is a beat reward that makes each additional beat count as a credit against the path cost. Without this term, the optimizer could favor short fragments that accumulate little penalty. The beat reward ensures long paths covering the full recording are competitive. 

The total cost of a candidate heartbeat sequence is defined as
\begin{equation}
    J(\hat{\mathcal{R}}) = \sum_{(i,j)\in\hat{\mathcal{R}}} C(i,j) 
    + C_{\text{endpoint}}(\hat{\mathcal{R}}) 
    + C_{\text{coverage}}(\hat{\mathcal{R}})
    \label{eq:total_cost}
\end{equation}
where endpoint penalties discourage paths that begin too late or end too early. Let $\hat{r}_1$ and $\hat{r}_M$ denote the first
and the final selected candidate peaks, respectively. Then

\begin{equation}
C_{\mathrm{endpoint}}(\hat{\mathcal{R}}) =
\eta_s \max(0,\, t(\hat{r}_1) - \tau_{\max}) +
\eta_e \max(0,\, T_{\text{rec}} - t(\hat{r}_M) - \tau_{\max}),
\label{eq:endpoint}
\end{equation}

where $\eta_s$ and $\eta_e$ are the start- and end-penalty weights, respectively.
The first term penalizes paths whose first detected beat occurs excessively late relative to the beginning of the recording. The second term penalizes paths that terminate too early before the end of the signal.

A coverage penalty is additionally used to discourage selecting
only a short, clean fragment of the ECG:
\begin{equation}
    C_{\text{coverage}}(\hat{\mathcal{R}})
    = \eta_c \left(1 - \frac{t(\hat{r}_M) - t(\hat{r}_1)}{T_{\text{rec}}}\right),
    \label{eq:coverage}
\end{equation}
where $\eta_c$ is the penalty weight. The quantity
$\frac{t(\hat{r}_M) - t(\hat{r}_1)}{T_{\text{rec}}}$ represents
the fraction of the ECG recording spanned by the estimated
heartbeat sequence. Sequences covering only a small temporal
portion of the signal therefore receive larger penalties.



Together, the endpoint and coverage penalties ensure that the
selected solution corresponds to a physiologically consistent
heartbeat sequence spanning the full ECG recording rather than
a short low-cost segment. The optimal estimated R-peak sequence
is then obtained by
\[
\hat{\mathcal{R}}^{*} = \arg\min_{\hat{\mathcal{R}}\,\in\,\mathcal{F}}\;J(\hat{\mathcal{R}}),
\]
where $J(\hat{\mathcal{R}})$ is as defined in~\eqnref{total_cost}.

\begin{algorithm}[t]
\caption{Beam-search DP path selection}
\label{alg:dp}
\begin{algorithmic}[1]
\Require $\mathcal{C}$, $\{q_j\}$, $\{\mathrm{Pred}[j]\}$, beam width $B$,
         window $W$, HR prior $\hat{I}(t)$, $\sigma_\mathrm{HR}$,
         $\boldsymbol{\lambda} = [\lambda_\text{local}, \lambda_\text{HR},
         \lambda_\text{node}]^\top$, $\lambda_b$, $\eta$
\Ensure R-peaks $\mathcal{R}^{*}$, IBIs $I^*$, heart rate $HR^*$
\State \textbf{Begin}
\State \textbf{Let} $\mathbf{c}(i,j) \triangleq [C_\text{local},\, C_\text{HR},\, C_\text{node}(j)]^\top$
       \Comment{see line 13}
\State $\mathcal{B}_j \leftarrow \emptyset$ for all $j$;\;
       initialise seed states at first reachable nodes
\For{each node $c_j$ in temporal order}
    \For{each $(i, w_{ij}) \in \mathrm{Pred}[j]$}
        \For{each state $S=(D,\pi,\Omega,\ell,s)\in\mathcal{B}_i$}
            \If{$|\Omega| \ge W_{\min}$}
                \State $C_\text{local} \leftarrow
                       \frac{|w_{ij}-\mathrm{med}(\Omega)|}
                            {\mathrm{MAD}(\Omega)+\epsilon}$
            \Else
                \State $C_\text{local}\leftarrow 0$ \Comment{insufficient history; suppress term}
            \EndIf
            \State $C_\text{HR} \leftarrow
                   |w_{ij}-\hat{I}(t_j)|\,/\,\sigma_\mathrm{HR}$
            \State $D_\text{new} \leftarrow D +
                   \boldsymbol{\lambda}^\top\mathbf{c}(i,j) - \lambda_b$
                   \State$\Omega_\text{new} \leftarrow
                   \mathrm{SlideWindow}(\Omega,w_{ij},W)$ 
            \State Insert $(D_\text{new},\,j,\,\Omega_\text{new},\,
                   \ell{+}1,\,s)$ into $\mathcal{B}_j$
        \EndFor
    \EndFor
    \State Prune $\mathcal{B}_j$ to best-$B$ states by $D$
\EndFor
\For{each terminal state $S\in\bigcup_j\mathcal{B}_j$}
    \State $J(S)\leftarrow D(S)
           + \eta\,C_\text{endpoint}(S)
           + \eta\,C_\text{coverage}(S)$
\EndFor
\State $S^*\leftarrow\arg\min_S J(S)$;\;
       backtrack parent pointers $\to R^*$
\State $I_k^*\leftarrow t(r_{k+1}^*)-t(r_k^*)$;\;
       $HR_k^*\leftarrow \gamma/I_k^*$
\State \Return $\mathcal{R}^{*},\;I^*,\;HR^*$ 
\State \textbf{End}
\end{algorithmic}
\end{algorithm}
\subsection{Beam-Search Dynamic Programming}
A standard shortest-path algorithm keeps one best cost per node. ARTA cannot do this because future transition costs depend on the path-specific IBI history $\Omega$. Two paths ending at the same candidate may have different rhythm histories and, therefore, different future costs. 

To preserve multiple plausible rhythm histories, the method uses beam-search dynamic programming (DP). For each node \(c_j\), the algorithm stores a beam
\[
B_j = \{S_{j,1}, S_{j,2}, \ldots, S_{j,B}\},
\]
where \(B\) is the beam width. Each beam state stores
\[
S_{j,b}
=
\left(
D_{j,b},
\pi_{j,b},
\Omega_{j,b},
\ell_{j,b}
\right),
\]
where \(D_{j,b}\) is the accumulated cost, \(\pi_{j,b}\) is the parent pointer, \(\Omega_{j,b}\) is the recent IBI window, and \(\ell_{j,b}\) is the path length.

For every incoming edge and every predecessor beam state, a new candidate state is generated as
\[
D_{\mathrm{new}}
=
D_{\mathrm{prev}}
+
C(i,j).
\]

The updated rhythm memory is 
\[
\Omega_{\mathrm{new}}
=
\operatorname{SlideWindow}
\left(
\Omega_{\mathrm{prev}},
w_{ij},
W
\right).
\]

Only the top \(B\) states are preserved at each node. This prevents premature pruning of potentially correct rhythm histories.

The candidate state is inserted into \(B_j\), which retains only the \(B\) lowest-cost states after each update. Endpoint and coverage penalties are applied to terminal states at the end of the signal. The best terminal state is selected, and parent pointers are backtracked to recover \(R^{*}\).~\algoref{dp} provides complete pseudocode. The computational complexity is \(O(LKB)\), where \(L\) is the number of candidates, \(K\) is the maximum in-degree per node (bounded by physiological RR limits), and \(B\) is the beam width, making the algorithm practical for typical recording lengths.

\begin{table}[t]
\caption{Summary of notation and parameter values used in all experiments.}
\label{tab:notation_value}
\centering
\begin{tabular}{lll}
\toprule
Symbol & Description & Value \\
\midrule
\multicolumn{3}{c}{\textit{Signal and evaluation}} \\
\midrule
$x[n]$                      & Noisy ECG signal, $n=1,\ldots,N$          & —         \\
$f_s$                       & Sampling frequency                        & 500 Hz    \\
$\mathcal{R}$               & True (unknown) R-peak sequence            & —         \\
$\hat{\mathcal{R}}$         & Estimated R-peak sequence                 & —         \\
$\hat{r}_k$                 & $k$-th estimated R-peak                   & —         \\
$\hat{I}_k$                 & $k$-th estimated IBI: $t(\hat{r}_{k+1})-t(\hat{r}_k)$ & — \\
$T_\text{rec}$              & Recording duration                        & —         \\
$T_w,\,T_s$                 & Evaluation window duration and stride     & 8 s, 2 s  \\
$\gamma$                    & Millisecond-to-BPM constant               & 60{,}000  \\
$\delta$                    & Beat-matching tolerance                   & 75 ms     \\
\midrule
\multicolumn{3}{c}{\textit{Candidate graph}} \\
\midrule
$\mathcal{C}$               & Over-complete candidate set               & —         \\
$c_j$                       & $j$-th candidate node                     & —         \\
$q_j,\,A_j$                 & Confidence score and local amplitude of $c_j$ & —     \\
$w_{ij}$                    & Interval between $c_i$ and $c_j$         & —         \\
$\mathrm{Pred}[j]$          & Valid predecessors of $c_j$               & —         \\
$\tau_{\min},\,\tau_{\max}$ & Admissible RR interval bounds             & 300, 2000 ms \\
\midrule
\multicolumn{3}{c}{\textit{Rhythm memory and prior}} \\
\midrule
$\Omega_i$                  & Path-local IBI window at node $c_i$      & —         \\
$W,\,W_{\min}$              & IBI window length and minimum length      & 5, 2 beats \\
$m_i,\,s_i$                 & Median and MAD of $\Omega_i$              & —         \\
$\hat{I}(t)$                & Soft time-varying HR prior (ms)           & —         \\
$\sigma_{\mathrm{HR}}$      & HR prior tolerance                        & —         \\
\midrule
\multicolumn{3}{c}{\textit{Cost function and optimization}} \\
\midrule
$C_\text{rhythm}$           & Path-local rhythm cost                    & —         \\
$C_\text{prior}$            & HR-prior agreement cost                   & —         \\
$C_\text{node}$             & Candidate confidence cost                 & —         \\
$\boldsymbol{\lambda}$      & Cost weight vector $[\lambda_\text{rhythm},\lambda_\text{prior},\lambda_\text{node}]^\top$ & [1.15, 0.75, 0.18] \\
$\lambda_b$                 & Beat reward                               & 0.42      \\
$[\eta_s,\eta_e,\eta_c]$    & Start, end, and coverage penalty weights  & [0.04, 0.04, 0.25] \\
$J(\hat{\mathcal{R}})$      & Total path objective                      & —         \\
$\mathcal{B}_j,\,B$         & Beam at node $c_j$ and beam width         & —, 3      \\
\bottomrule
\end{tabular}
\end{table}
\section{Experimental Setup}
\label{sec:experiment}

\subsection{Datasets, Preprocessing, and Evaluation Design}
We evaluated ARTA using two complementary ECG domains. The IEEE Signal Processing Cup 2015 dataset (SPC 2015) served as the primary controlled exercise benchmark, representing young healthy participants undergoing treadmill running with known window-level heart-rate references. The European ST--T Database (EDB) was used as a cross-dataset clinical stress test. EDB introduces long-duration ambulatory recordings, heterogeneous lead configurations, abnormal ST--T morphology, rhythm variation, ectopic beats, and naturally occurring signal-quality changes. This two-domain design allowed us to evaluate both robustness within the controlled exercise setting and generalization beyond that setting. No PPG, accelerometer, or EDB ST--T episode annotations were used by ARTA.

\subsubsection{IEEE Signal Processing Cup 2015 Dataset}

The primary experiments used ECG recordings from the IEEE Signal Processing Cup 2015 dataset (SPC 2015), a widely adopted benchmark for cardiac signal processing during exercise and motion~\cite{zhang2014troika}. The dataset contains simultaneous two-channel PPG, three-axis accelerometer, and single-channel ECG recordings from 12 healthy participants aged 18–35, acquired at 125~Hz during treadmill running at varying speeds. Only the ECG channel was used in this study; PPG and accelerometer measurements were excluded from candidate generation, graph construction, inference, and evaluation, consistent with the ECG-only design of ARTA.

The ECG was resampled from 125 to 500~Hz using polyphase resampling to standardize the computational sampling grid used for candidate localization and graph inference, following the convention of prior work on this dataset~\cite{Aygun,huang2021graph}. The corresponding BPM trace files were used as window-level ground-truth heart-rate references. The BPM0 labels are defined over $8$~s windows with $ 2$~s stride, and the same window structure was applied during evaluation so that BPM estimates derived from detected R-peaks could be directly compared against the reference labels.

\subsubsection{European ST--T Database}

Cross-dataset evaluation was performed using the European ST--T Database (EDB), a clinical ambulatory ECG dataset containing 90 two-hour recordings from 79 subjects with diagnosed or suspected myocardial ischemia~\cite{taddei1992european}. Each record contains two ECG leads sampled at 250~Hz and expert-reviewed beat annotations. EDB was selected because it challenges several assumptions of the primary IEEE SPC 2015 benchmark: its recordings are substantially longer with lower average heart rates, its lead configurations vary between records, and its clinical population exhibits abnormal ST--T morphology, ectopy, rhythm variation, and native ambulatory signal-quality changes. EDB therefore provides a stringent test of whether ARTA generalizes beyond healthy subjects under controlled exercise.

A fixed, label-independent lead-selection rule was used: the first listed ECG lead from each record was analyzed. This avoids selecting the best lead using reference annotations while retaining the lead heterogeneity of the database. The selected ECG was resampled from 250 to 500~Hz using polyphase resampling. Any non-finite samples were linearly interpolated before resampling. Expert beat annotations were mapped to the 500-Hz sample grid and used as the beat-level reference. Window-level reference heart rate was derived from annotated RR intervals over 8-s windows with a 2-s stride, matching the primary evaluation protocol.

The unmodified EDB ECG was further corrupted using controlled electrode-motion artifact from the MIT--BIH Noise Stress Test Database. The use of synthetic artifact in addition to native ambulatory variability provided known corruption conditions while preserving the EDB beat annotations. Six nominal SNR conditions, $-2$, 0, 2, 4, 6, and 12~dB, were evaluated. Because every EDB record is two hours long, inference was performed independently over consecutive non-overlapping 120-s segments. Only evaluation windows fully contained within a segment were scored. Segment-level results were then aggregated over all records and corruption conditions. ARTA used beam width $B=3$ and local rhythm memory $\Omega=3$ throughout this evaluation. A sensitivity analysis showed that increasing the rhythm memory to $\Omega=5$ did not improve EDB performance; therefore, the shorter $\Omega=3$ memory was retained because it adapted more quickly to nonstationary rhythm and candidate corruption. All complete EDB segments were retained in the primary cross-dataset analysis. No signal-quality or outcome-based exclusion was applied to the primary result. A separate post-hoc quality-based sensitivity analysis examined performance after selectively retaining high-quality segments using only label-free ARTA diagnostics; reference annotations were not used to determine segment retention.

\subsubsection{Motion-Artifact Generation}

Controlled electrode-motion corruption was generated using the \texttt{em} record from the MIT--BIH Noise Stress Test Database (NSTDB)~\cite{Moody2001TheIO}. Electrode-motion artifact was selected because it can introduce QRS-like transients, baseline shifts, missed beats, and false detections, making it particularly challenging for IBI estimation. The noise record was resampled to 500~Hz, and random starting offsets were used to obtain artifact excerpts. For signals longer than the available artifact recording, the noise was extended to the required duration while maintaining temporal alignment between the source ECG, corrupted ECG, and reference labels.

For target SNR $s$, the corrupted ECG was generated as

\begin{equation}
x_{\mathrm{noisy}}[n]
=
x_{\mathrm{source}}[n]
+
\alpha_s n_{\mathrm{EM}}[n],
\end{equation}

where $n_{\mathrm{EM}}[n]$ denotes zero-mean electrode-motion artifact and

\begin{equation}
\alpha_s
=
\sqrt{
\frac{P_x}
     {P_n\,10^{s/10}}
},
\qquad
P_x =
\frac{1}{N}\sum_{n=1}^{N}
\left(x_{\mathrm{source}}[n]-\bar{x}_{\mathrm{source}}\right)^2,
\qquad
P_n =
\frac{1}{N}\sum_{n=1}^{N}
\left(n_{\mathrm{EM}}[n]-\bar{n}_{\mathrm{EM}}\right)^2.
\end{equation}

Both datasets were evaluated at six target SNR levels: $-2$, 0, 2, 4, 6, and 12~dB. Within each dataset and SNR condition, the same noisy ECG inputs were provided to every compared method to ensure a paired and consistent comparison.

\subsubsection{Label-Free Selective-Reliability Analysis}
\label{sec:selective_reliability}

To examine whether ARTA can identify unreliable ECG portions without access to reference annotations, we performed a post-hoc selective-reliability analysis on the 120-s EDB segment instances. Each segment was assigned an ARTA-derived reliability score using four label-free internal diagnostics: mean local candidate signal quality, multi-detector consensus fraction, mean confidence in the time-varying HR prior, and the temporal span covered by the selected beat path. These quantities were calculated entirely from the noisy ECG, candidate detections, and selected ARTA path; reference R-peaks, BPM labels, and estimation errors were not used to construct the score. Each diagnostic was robustly normalized to $[0,1]$ using its pooled EDB 1st and 99th percentiles, and the four normalized values were averaged with equal weight. Complete mathematical definitions of the diagnostics, normalization bounds, and reliability score are provided in Appendix~\ref{app:arta_reliability}.

All 32,400 record--SNR segment instances were ranked globally by their reliability scores. Performance was recomputed after retaining the highest-scoring 100\%, 90\%, 80\%, 70\%, 60\%, 50\%, 40\%, and 30\% of segments. The 100\% operating point, which includes every complete segment, remains the primary EDB result. The lower-retention operating points quantify the conditional performance--coverage trade-off and do not represent a prospectively calibrated rejection threshold.

\subsection{Implementation and Parameter Settings}
All experiments were implemented in Python using NumPy and SciPy for signal processing, graph construction, dynamic programming, and numerical evaluation. ECG recordings were first resampled to $f_s = 500$~Hz, and all candidate detection, IBI estimation, and evaluation steps were performed at this sampling frequency. The proposed method operates on an over-complete ECG R-peak candidate set generated from multiple ECG-only detectors. These candidates are represented as nodes in a directed acyclic graph (DAG), where an edge is admitted only when the corresponding RR interval lies within the physiologically admissible range $\tau_{\min} = 300$~ms to $\tau_{\max} = 2000$~ms, preventing the graph search from selecting paths with unrealistically short or long interbeat intervals.

The final R-peak sequence is estimated by minimizing a weighted path objective that combines local rhythm consistency, agreement with a soft time-varying HR prior, and candidate-node confidence. 
The HR prior $\hat{I}(t)$ is estimated from the ECG itself using overlapping windows of duration $T_w = 8$~s with stride $T_s = 2$~s and is converted to a time-varying expected IBI. A tolerance parameter of $\sigma_{\mathrm{HR}} = 200$~ms is used for the HR-prior penalty, allowing the prior to guide the path search without enforcing a fixed heart rate.

The cost weights were fixed across all experiments as $\lambda_{\mathrm{rhythm}} = 1.15$, $\lambda_{\mathrm{prior}} = 0.75$, and $\lambda_{\mathrm{node}} = 0.18$, such that local rhythm consistency exerts the strongest influence, the HR prior acts as a softer physiological constraint, and candidate confidence serves primarily as a secondary tie-breaking term. A beat reward of $\lambda_b = 0.42$ encourages selection of complete beat trajectories rather than overly sparse paths. Start, end, and coverage penalty weights were set to $\eta_s = 0.04$, $\eta_e = 0.04$, and $\eta_c = 0.25$, respectively, to discourage solutions that explain only a short clean portion of the recording. Dynamic programming was performed with beam search using a beam width of $B = 3$ states per graph node. The path-local rhythm-memory length was set to $W=5$ for the SPC 2015
experiments. For EDB, $W=3$ was selected using the fixed development-subset
sensitivity analysis reported in Appendix~\ref{app:edb_memory}. Beam width and
all cost weights were held fixed across records and SNR conditions within each
dataset.
~\tblref{notation_value} summarizes all notation and parameter values.

\subsection{Evaluation Metric}
Performance was evaluated at both the window level and the beat level in order to assess heart-rate accuracy, rhythm consistency, and R-peak localization performance under motion artifact.
\subsubsection{Window-Level HR Metrics}
Evaluation at the window level operates on estimated BPM values \(\hat{h}_i\) compared against reference values \(h_i\) over \(K\) total windows.

\textbf{Mean Absolute Error (MAE) } is defined as 
\small{
\[
\small
\mathrm{MAE}_{\mathrm{BPM}}
=
\frac{1}{N}
\sum_{i=1}^{N}
\left|
\hat{h}_i - h_i
\right|.
\]
}
MAE measures the average magnitude of BPM estimation error and is less sensitive to occasional outliers.

\textbf{Root Mean Square Error (RMSE)} squares the residuals before averaging, making it sensitive to large individual errors:

\[
\small
\mathrm{RMSE}_{\mathrm{BPM}}
=
\sqrt{
\frac{1}{N}
\sum_{i=1}^{N}
\left(
\hat{h}_i - h_i
\right)^2
}
\]


\textbf{Pearson Correlation Coefficient} ($r$) measures whether estimated and reference HR track each other across the dynamic range of the recording, independent of any fixed offset or scale:
\[
r
=
\frac{
\sum_{i=1}^{N}
\left(
\hat{h}_i - \bar{\hat{h}}
\right)
\left(
h_i - \bar{h}
\right)
}{
\sqrt{\sum_{i=1}^{N} \left(\hat{h}_i - \bar{\hat{h}}\right)^2}
\sqrt{
\sum_{i=1}^{N}
\left(
h_i - \bar{h}
\right)^2
}
}
\]
This is particularly relevant during exercise, where HR changes substantially over time. A method could have low MAE on a flat resting segment while completely failing to track a tachycardic episode; $r$ captures this. 

\textbf{Tolerance-based accuracy} was evaluated using the percentage of windows satisfying
\(
\left|
\hat{h}_i - h_i
\right|
\leq 5 \ \mathrm{BPM}.
\). 
\[
W_{\pm 5}
=
\frac{1}{N}
\sum_{i=1}^{N}
\mathbf{1}
\left[
\left|
\hat{h}_i - h_i
\right|
\leq 5
\right].
\]

\subsubsection{Beat-Level Metrics}
Beat-level metrics are evaluated by matching detected peaks \(\hat{\mathcal{R}}\) against clean-ECG derived reference peaks \(\mathcal{R}^{*}\) within a timing tolerance of \(\delta = 75\,\mathrm{ms}\). A detected peak is considered a true positive (TP) if a reference peak exists within \(\delta\); otherwise, it is counted as a false positive (FP). A reference peak with no corresponding match within \(\delta\) is counted as a false negative (FN).

Standard detection metrics including \textbf{precision}, \textbf{recall}, and \textbf{F1-score} are then computed from the TP, FP, and FN counts. Precision measures the proportion of detected peaks that correspond to true heartbeats, while recall measures the proportion of reference heartbeats successfully recovered by the algorithm. The F1-score provides a balanced summary of detection performance by jointly considering false positives and missed beats.

\textbf{IBI MAE} measures the accuracy of interval duration estimates over all \(M-1\) consecutive matched beat pairs:
\[
\mathrm{IBI\text{-}MAE}
=
\frac{1}{M-1}\sum_{k=1}^{M-1}
\left| \hat{I}_k - I_k^{*} \right|,
\]
where \(\small \hat{I}_k = t(\hat{r}_{k+1}) - t(\hat{r}_k)\)
and \(\small I_k^{*} = t(r_{k+1}^{*}) - t(r_k^{*})\). IBI MAE is reported in milliseconds and provides the most direct measure of HRV-relevant accuracy, since downstream HRV indices such as SDNN and RMSSD are computed directly from the IBI sequence.

\textbf{Beat Timing Error} measures the mean absolute temporal displacement of each matched detected peak from its reference:
\[
\mathrm{Timing\ Error}=
\frac{1}{\mathrm{M}}\sum_{\mathrm{M}}
\left| t(\hat{r}) - t(r^{*})\right|,
\]

This metric is distinct from IBI MAE. Two consecutive peaks may each be shifted by the same amount, producing zero IBI error but nonzero timing error. Conversely, alternating early and late detections can inflate timing error while partially cancelling in the IBI sequence.

\begin{table*}[t]
\centering
\caption{Overall state-of-the-art comparison across both datasets.}
\label{tab:overall_comparison}
\begin{tabular}{l cccc cccc}
\toprule
& \multicolumn{4}{c}{IEEE Signal Cup (IEEE SPC 2015)} & \multicolumn{4}{c}{European ST--T (EDB)} \\
\cmidrule(lr){2-5} \cmidrule(lr){6-9}
Method & MAE BPM & RMSE BPM & Pearson \(r\) & \(W_{\pm 5}\)
       & MAE BPM & RMSE BPM & Pearson \(r\) & \(W_{\pm 5}\) \\
\midrule
ARTA                      & \textbf{1.796} & \textbf{5.212} & \textbf{0.977} & \textbf{92.38}\%
                          & \textbf{10.084} & \textbf{18.956} & \textbf{0.54} & \textbf{62.28}\% \\
Aygun et al.~\cite{Aygun} & 14.858 & 19.889 & 0.745 & 28.34\%
                          & 25.594 & 34.924 & 0.238 & 31.52\% \\
WFDB--GQRS                & 9.765 & 23.847 & 0.451 & 76.73\%
                          & 53.697 & 63.711 & 0.122 & 16.50\% \\
Tiramisu~\cite{tiramisu}  & 5.787 & 13.021 & 0.853 & 72.87\%
                          & 42.803 & 49.287 & 0.045 & 10.28\% \\
\midrule
WFDB--XQRS                & 3.929 & 10.275 & 0.920 & 84.18\%
                          & 40.060 & 52.141 & 0.225 & 30.43\% \\
Pan--Tompkins             & 6.196 & 11.478 & 0.893 & 66.73\%
                          & 32.981 & 41.095 & 0.165 & 20.93\% \\
Wavelet                   & 11.275 & 16.646 & 0.731 & 41.73\%
                          & 49.206 & 55.047 & -0.002 & 6.96\% \\
Majority (2/3)            & 6.045 & 11.838 & 0.878 & 70.11\%
                          & 37.900 & 47.218 & 0.151 & 20.14\% \\
Union-NMS                 & 63.068 & 79.565 & 0.454 & 19.65\%
                          & 111.230 & 125.070 & 0.025 & 4.44\% \\
\bottomrule
\end{tabular}
\end{table*}

\begin{table*}[t]
\centering
\caption{Overall derived beat-level metrics comparison across both datasets.}
\label{tab:beat_level_metrics}
\begin{tabular}{l ccccc ccccc}
\toprule
& \multicolumn{5}{c}{IEEE Signal Cup (IEEE SPC 2015)} & \multicolumn{5}{c}{European ST--T (EDB)} \\
\cmidrule(lr){2-6} \cmidrule(lr){7-11}
Method & F1 & Pre & Rec & IBI MAE & Timing MAE
       & F1 & Pre & Rec & IBI MAE & Timing MAE \\
\midrule
ARTA                      & \textbf{0.95} & \textbf{0.946} & 0.95 & \textbf{6.52} & 10.860
                          & \textbf{0.689} & \textbf{0.665} & 0.715 & 18.368 & 49.010 \\
Aygun et al.~\cite{Aygun} & 0.543 & 0.591 & 0.502 & 16.926 & 14.482
                          & 0.450 & 0.437 & 0.463 & 21.784 & 50.805 \\
WFDB--GQRS                & 0.916 & 0.889 & 0.944 & 12.942 & 36.162
                          & 0.594 & 0.482 & 0.775 & 21.387 & \textbf{26.096} \\
Tiramisu~\cite{tiramisu}  & 0.910 & 0.906 & 0.913 & 7.850 & 14.575
                          & 0.577 & 0.495 & 0.690 & \textbf{13.235} & 44.550 \\
\midrule
WFDB--XQRS                & 0.941 & 0.930 & \textbf{0.953} & 10.794 & 9.095
                          & 0.690 & 0.596 & 0.820 & 20.381 & 46.590 \\
Pan--Tompkins             & 0.885 & 0.914 & 0.859 & 13.125 & \textbf{8.359}
                          & 0.574 & 0.517 & 0.644 & 19.691 & 49.631 \\
Wavelet                   & 0.735 & 0.762 & 0.710 & 13.411 & 10.063
                          & 0.490 & 0.416 & 0.594 & 22.653 & 49.944 \\
Majority (2/3)            & 0.895 & 0.908 & 0.882 & 12.163 & 7.545
                          & 0.580 & 0.515 & 0.663 & 20.389 & 49.557 \\
Union-NMS                 & 0.866 & 0.778 & \textbf{0.978} & 11.983 & 7.124
                          & 0.582 & 0.445 & \textbf{0.841} & 25.406 & 48.449 \\
\bottomrule
\end{tabular}
\end{table*}

\section{Results}

We evaluate ARTA at both the window level (heart-rate accuracy) and the beat level (R-peak localization and IBI fidelity), comparing against state-of-the-art methods - Aygun et al.~\cite{Aygun} and the Tiramisu denoising model~\cite{tiramisu} - as well as the individual ECG detectors that form our candidate set (WFDB-XQRS, Pan-Tompkins, wavelet and WFDB-GQRS) and two naive fusion baselines (Majority voting and Union-NMS). All experiments use the SNR levels $-2$, 0, 2, 4, 6, and 12~dB, with parameters held fixed across every subject and noise condition. The overall performance tables (~\tblref{overall_comparison} and~\tblref{beat_level_metrics}) report results for both datasets. Unless explicitly identified as an EDB analysis, the subsequent SNR-stratified, HRV, subject-level, ablation, and computational evaluations use SPC 2015.

\subsection{Overall Performance Comparison}
Tables~\ref{tab:overall_comparison} and~\ref{tab:beat_level_metrics}
summarize the primary full-coverage results on the controlled treadmill
dataset and the cross-dataset EDB cohort. No reliability-based segment
exclusion was applied to the results in either table. The two datasets
represent complementary evaluation domains: the treadmill dataset measures
robustness under controlled exercise-related corruption, whereas EDB tests
transfer to long clinical ambulatory ECG with heterogeneous leads, abnormal
morphology, rhythm variation, and native signal-quality changes.

On the SPC 2015 dataset (~\tblref{overall_comparison}), ARTA achieved $\mathrm{MAE}=1.796$~BPM and $W_{\pm5}=92.38\%$, corresponding to a 54.3\% reduction in MAE and an 8.2\% improvement in $W_{\pm5}$ relative to the strongest single-detector baseline, WFDB--XQRS. Pan--Tompkins, Tiramisu, and the original graph-based method of Aygun~\textit{et al.} produced MAEs equal to $3.45\times$, $3.22\times$, and $8.27\times$ that of ARTA, respectively. These differences indicate that graph-constrained sequence selection is more robust than relying on an individual detector or an average-HR-dependent graph prior under sustained motion artifact.


On EDB, absolute accuracy decreased for all evaluated methods, reflecting the substantial shift from controlled exercise ECG to long clinical ambulatory recordings. Nevertheless, ARTA remained the strongest method for window-level HR estimation. Relative to the strongest competing method by MAE, Aygun et al., ARTA reduced MAE by 60.6\% and RMSE by 45.7\%, while increasing $W_{\pm5}$ by 30.76 percentage points. ARTA also produced substantially stronger correlation with the reference HR trajectory. Thus, although the clinical domain shift increased the absolute difficulty of the task, it did not eliminate the comparative advantage of graph-constrained temporal selection.

On SPC 2015, two fusion baselines included as analytical controls reveal something important about why graph-based selection is necessary. Majority voting---accepting a candidate only when at least two of the three detectors agree---achieves $\mathrm{MAE} = 6.045$~BPM and $W_{\pm 5} = 70.11\%$, which is $3.37\times$ worse in MAE than ARTA despite using the same three detectors as inputs. Union-NMS, which accepts every candidate from any detector, produces $\mathrm{MAE} = 63.068$~BPM, a $35\times$ degradation relative to ARTA. The contrast between these two extremes is instructive: being maximally selective (voting) discards valid auxiliary candidates that only one detector found, while being maximally permissive (union) floods the IBI sequence with false positives. Neither strategy has access to the temporal rhythm context that distinguishes a true missed beat from an artifact, which is precisely what the graph-based path optimization provides.

The beat-level results in Table~\ref{tab:beat_level_metrics} reinforce the
window-level findings. On the SPC 2015 dataset, ARTA achieved
$\mathrm{F1}=0.950$, with precision of 0.946 and recall of 0.950. Its balanced
precision and recall, together with improvements over Majority voting in both
quantities, are consistent with the rhythm-constrained path simultaneously
suppressing spurious candidates and recovering beats missed by individual
detectors. Although Union-NMS achieved the highest recall through permissive
candidate acceptance, its poor BPM accuracy demonstrates that high recall
alone does not guarantee a physiologically coherent IBI sequence.

At the beat level on EDB in Table~\ref{tab:beat_level_metrics} , ARTA achieved an F1 score of 0.679, precision of 0.682, and recall of 0.675. Its precision was the highest among the evaluated methods, indicating stronger rejection of artifact-induced candidate peaks. WFDB--XQRS obtained a slightly higher F1 score and recall, while Tiramisu obtained a lower conditional IBI MAE. Nevertheless, ARTA produced substantially better window-level HR estimates, indicating that its selected beats formed a more coherent temporal trajectory even when it did not maximize every individual beat-level metric. The lower absolute beat-level performance compared with the treadmill dataset reflects the greater morphological heterogeneity, ectopy, lead variability, and native signal-quality changes present in EDB.

ARTA's lower absolute performance on EDB reflects the substantially larger domain shift relative to SPC 2015. EDB contains long clinical ambulatory recordings with heterogeneous lead configurations, abnormal ECG morphology, ectopic and irregular beats, and native signal-quality variation, in addition to the injected electrode-motion artifact. These conditions increase detector disagreement, reduce the reliability of the candidate pool, and can destabilize the ECG-derived HR prior. Moreover, ARTA can recover a true beat only when at least one constituent detector provides a plausible nearby candidate.

For EDB, the shorter rhythm-memory length was selected because it was less likely to preserve an incorrect rhythm after false candidates entered the graph. On the fixed development subset, using $\Omega=3$ instead of $\Omega=5$ reduced BPM MAE and RMSE by 14.8\% and 14.2\%, respectively; the complete sensitivity analysis is provided in Appendix~\ref{app:edb_memory}. The primary EDB results in Tables~\ref{tab:overall_comparison} and~\ref{tab:beat_level_metrics} nevertheless include all complete segments without reliability-based exclusion. This performance gap therefore reflects the broader clinical domain shift rather than rhythm-memory length alone and motivates the subsequent label-free reliability analysis.

\subsection{Label-Free Selective-Reliability Analysis}
\label{sec:edb_reliability}

We examined whether ARTA could identify less reliable EDB outputs using only information available during inference. Each 120-s segment--SNR instance was assigned a label-free reliability score combining local candidate signal quality, detector consensus, HR-prior confidence, and selected-path temporal span. Neither reference R-peaks nor BPM labels were used to calculate this score or rank the segments. The diagnostic definitions and score construction are provided in Appendix~\ref{app:arta_reliability}, with additional reliability results reported in Appendix~\ref{app:arta_reliability_results}. The primary EDB evaluation retained all segments (reported in
~\tblref{overall_comparison} and~\tblref{beat_level_metrics}); the following analysis was performed separately as a post-hoc performance--coverage sensitivity study.

\begin{figure}[]
    \centering
    \includegraphics[width=0.88\textwidth]
    {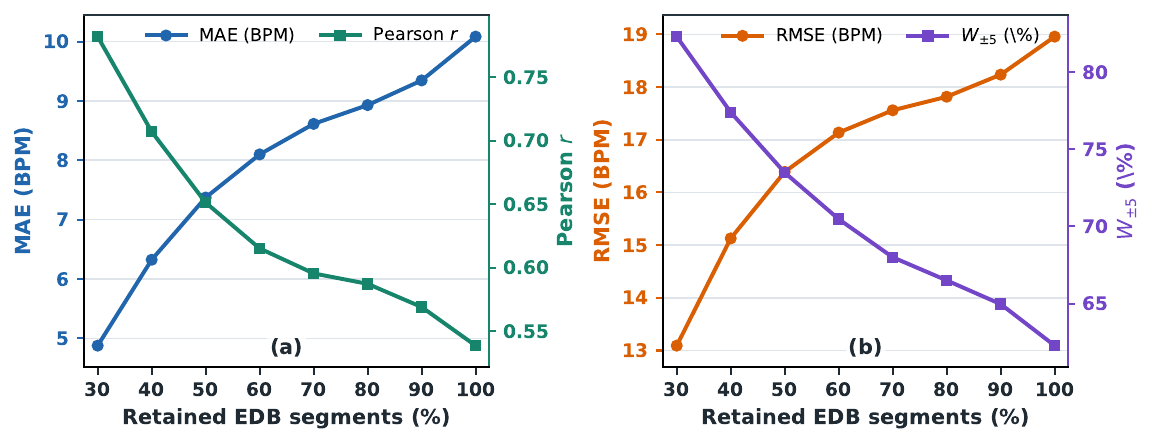}
    \caption{Label-free performance--coverage sensitivity on EDB.
    (a) BPM MAE (left axis) and Pearson correlation (right axis).
    (b) BPM RMSE (left axis) and percentage of windows within
    $\pm5$~BPM (right axis). Metrics were recomputed after retaining
    progressively larger proportions of the highest-scoring segment--SNR
    instances. The 100\% point represents the primary unfiltered EDB
    evaluation.}
    \label{fig:edb_quality_tradeoff}
\end{figure}

~\figref{edb_quality_tradeoff} shows that the reliability score was negatively associated with segment-level BPM MAE (Spearman $\rho=-0.496$) and positively associated with R-peak F1 ($\rho=0.49$). At 80\% segment retention, BPM MAE decreased from 10.1 to 8.93~BPM, Pearson correlation increased from 0.539 to 0.587, and $W_{\pm5}$ increased from 62.3\% to 66.5\%. Beat-level performance showed the same trend: R-peak F1 increased from 0.68 to 0.72, while IBI MAE decreased from 18.4 to 15.9~ms. More restrictive operating points produced further improvements but reduced the amount of evaluated data. For example, at 50\% segment retention, BPM MAE was 7.37~BPM, R-peak F1 was 0.78, and IBI MAE was 11.5~ms. These results demonstrate a systematic conditional performance--coverage trade-off rather than an improvement to the primary full-coverage result.


These findings indicate that ARTA's internal diagnostics provide useful information about the reliability of its estimates. However, the continued improvement under more aggressive filtering is accompanied by a corresponding loss of segment coverage. The results therefore demonstrate an exploratory performance--coverage relationship rather than a prospectively calibrated segment-rejection rule.

\begin{figure*}   
    \centering
    \includegraphics[width=1\linewidth, trim={350 50 400 27},clip]{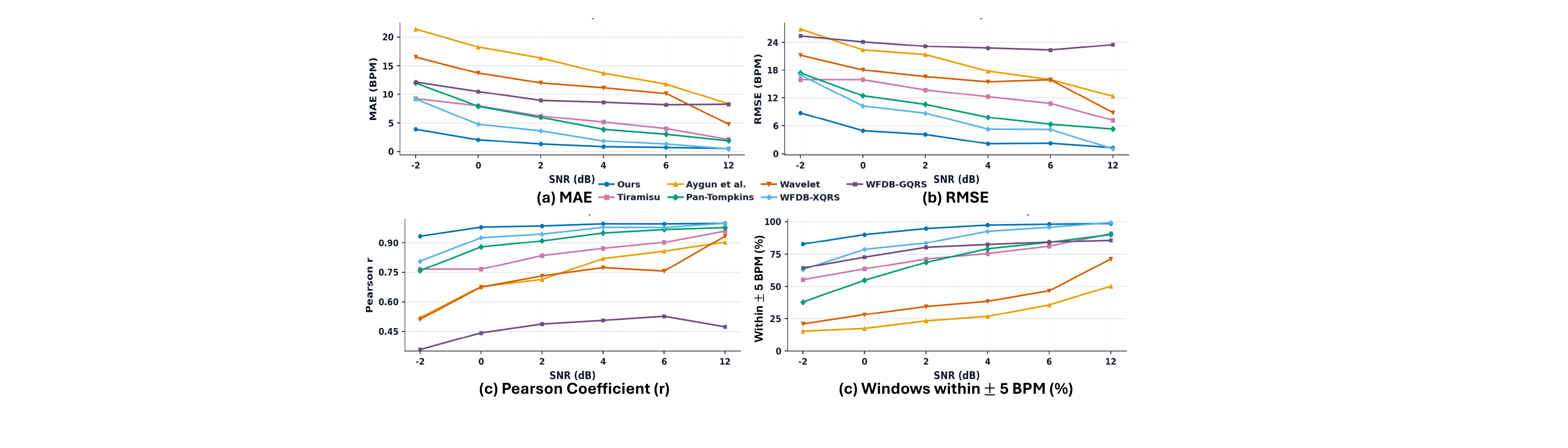}
    \caption{Window-level HR metrics versus SNR on SPC 2015: (a) MAE, (b) RMSE, (c) Pearson $r$ and (d) percentage of windows within $\pm 5$~BPM. ARTA (blue) is most robust across the full noise range.}
    \label{fig:snr_full}
\end{figure*}

\begin{table*}[t]
\caption{Pearson correlation between estimated and clean-ECG-derived HRV parameters on SPC 2015
         across SNR levels. Bold = best per row.}
\label{tab:hrv_pearson}
\centering
\setlength{\tabcolsep}{4pt}
\renewcommand{\arraystretch}{0.95}
\begin{tabular}{llccccccc}
\toprule
SNR & HRV Param.
    & \textbf{Ours}
    & XQRS
    & Tiramisu
    & Aygun
    & PT
    & Wavelet
    & GQRS \\
\midrule
\multirow{8}{*}{\rotatebox[origin=c]{90}{\small Mean}}
& Mean RR  & \textbf{0.992} & 0.995 & 0.925 & 0.469 & 0.596 & 0.609 & 0.920 \\
& SDNN     & 0.971 & \textbf{0.978} & 0.800 & 0.262 & 0.031 & 0.256 & 0.345 \\
& Mean HR  & 0.995 & \textbf{0.996} & 0.957 & 0.591 & 0.795 & 0.773 & 0.950 \\
& STD HR   & \textbf{0.931} & \textbf{0.928} & 0.661 & 0.281 & 0.318 & 0.327 & 0.433 \\
& VLF      & \textbf{0.950} & \textbf{0.950} & 0.490 & 0.163 & 0.024 & 0.180 & 0.441 \\
& LF       & 0.444 & \textbf{0.649} & 0.189 & 0.045 &-0.165 & 0.104 & 0.056 \\
& HF       & \textbf{0.347} & 0.251 &-0.217 & 0.564 & 0.216 & 0.424 & 0.065 \\
& Total    & 0.588 & \textbf{0.714} & 0.359 & 0.077 &-0.308 & 0.167 & 0.209 \\
\midrule
\multirow{8}{*}{\rotatebox[origin=c]{90}{\small 6 dB}}
& Mean RR  & \textbf{0.999} & 0.998 & 0.939 & 0.712 & 0.787 & 0.584 & 0.948 \\
& SDNN     & \textbf{0.998} & 0.997 & 0.920 & 0.515 & 0.156 & 0.223 & 0.393 \\
& Mean HR  & \textbf{1.000} & 0.999 & 0.965 & 0.801 & 0.906 & 0.785 & 0.973 \\
& STD HR   & \textbf{0.997} & 0.975 & 0.842 & 0.409 & 0.489 & 0.452 & 0.473 \\
& VLF      & \textbf{0.996} & 0.986 & 0.334 & 0.094 & 0.363 &-0.117 & 0.553 \\
& LF       & 0.904 & \textbf{0.963} & 0.558 &-0.055 &-0.302 & 0.073 &-0.121 \\
& HF       & 0.405 & \textbf{0.796} &-0.215 & 0.397 & 0.257 & 0.666 & 0.102 \\
& Total    & 0.858 & \textbf{0.964} & 0.539 &-0.045 &-0.415 & 0.129 & 0.340 \\
\midrule
\multirow{8}{*}{\rotatebox[origin=c]{90}{\small 2 dB}}
& Mean RR  & \textbf{0.987} & 0.994 & 0.908 & 0.475 & 0.429 & 0.558 & 0.903 \\
& SDNN     & \textbf{0.945} & \textbf{0.977} & 0.725 & 0.444 &-0.035 & 0.119 & 0.247 \\
& Mean HR  & 0.994 & \textbf{0.996} & 0.947 & 0.594 & 0.729 & 0.778 & 0.953 \\
& STD HR   & \textbf{0.830} & 0.923 & 0.509 & 0.459 & 0.279 & 0.348 & 0.367 \\
& VLF      & \textbf{0.936} & 0.984 & 0.630 & 0.333 &-0.113 & 0.166 & 0.608 \\
& LF       & 0.335 & 0.636 & -0.080& \textbf{0.694} &-0.196 &-0.056 & 0.336 \\
& HF       & 0.151 & -0.088& -0.317& \textbf{0.700} & 0.077 & 0.415 &-0.044 \\
& Total    & \textbf{0.414} & 0.553 & 0.357 & 0.338 &-0.300 & 0.042 & 0.168 \\
\midrule
\multirow{8}{*}{\rotatebox[origin=c]{90}{\small -2 dB}}
& Mean RR  & 0.982 & \textbf{0.985} & 0.866 &-0.110 & 0.361 & 0.537 & 0.876 \\
& SDNN     & \textbf{0.942} & 0.924 & 0.561 &-0.208 &-0.054 & 0.342 & 0.411 \\
& Mean HR  & 0.989 & \textbf{0.989} & 0.929 & 0.026 & 0.585 & 0.662 & 0.903 \\
& STD HR   & \textbf{0.890} & 0.822 & 0.431 &-0.005 & 0.109 & 0.116 & 0.414 \\
& VLF      & \textbf{0.806} & 0.754 & 0.105 &-0.230 &-0.291 & 0.561 &-0.037 \\
& LF       & -0.143& 0.129 &-0.250 &-0.009 &-0.147 & \textbf{0.426} &-0.312 \\
& HF       & 0.173 &-0.215 &-0.173 & \textbf{0.927} &-0.049 & 0.165 & 0.106 \\
& Total    & \textbf{0.596} & 0.503 &-0.124 & 0.004 &-0.313 & 0.455 &-0.043 \\
\bottomrule
\end{tabular}
\end{table*}
\begin{figure}[]
    \vspace{-1mm}
    \centering
    \includegraphics[width=0.9\linewidth, trim={5 8 9 22},clip]{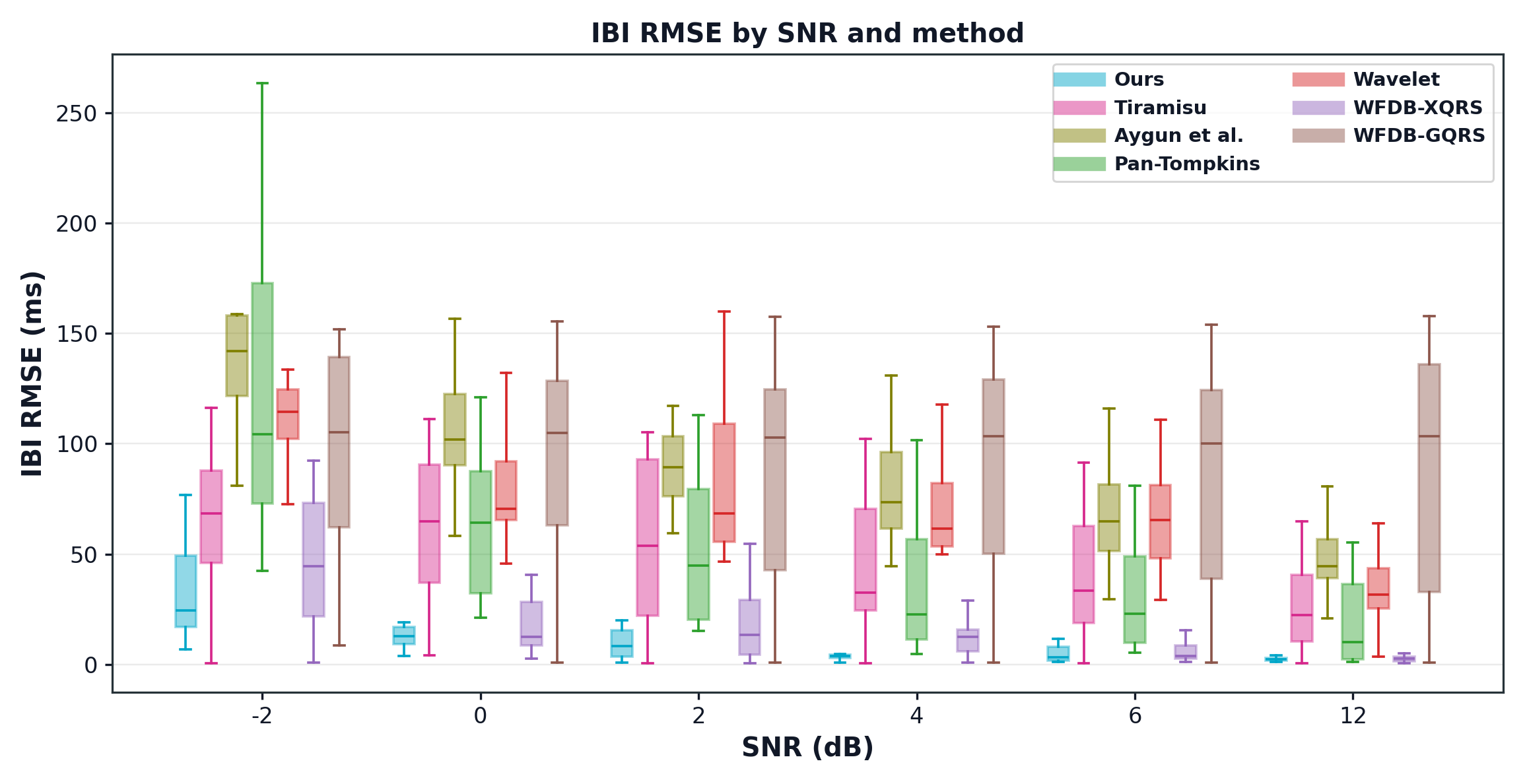}
    \vspace{-2mm}
    \caption{Distribution of beat-level IBI RMSE across SNR levels on SPC 2015. Each box summarizes subject-level variability at a given noise condition; ARTA (cyan) maintains the lowest median and tightest interquartile range across the full range.}
    \label{fig:snr_boxplot_IBI}
    \vspace{-2mm}
\end{figure}

\begin{figure}
    \vspace{-1mm}
    \centering
    \includegraphics[width=0.75\linewidth, trim={5 7 20 21},clip]{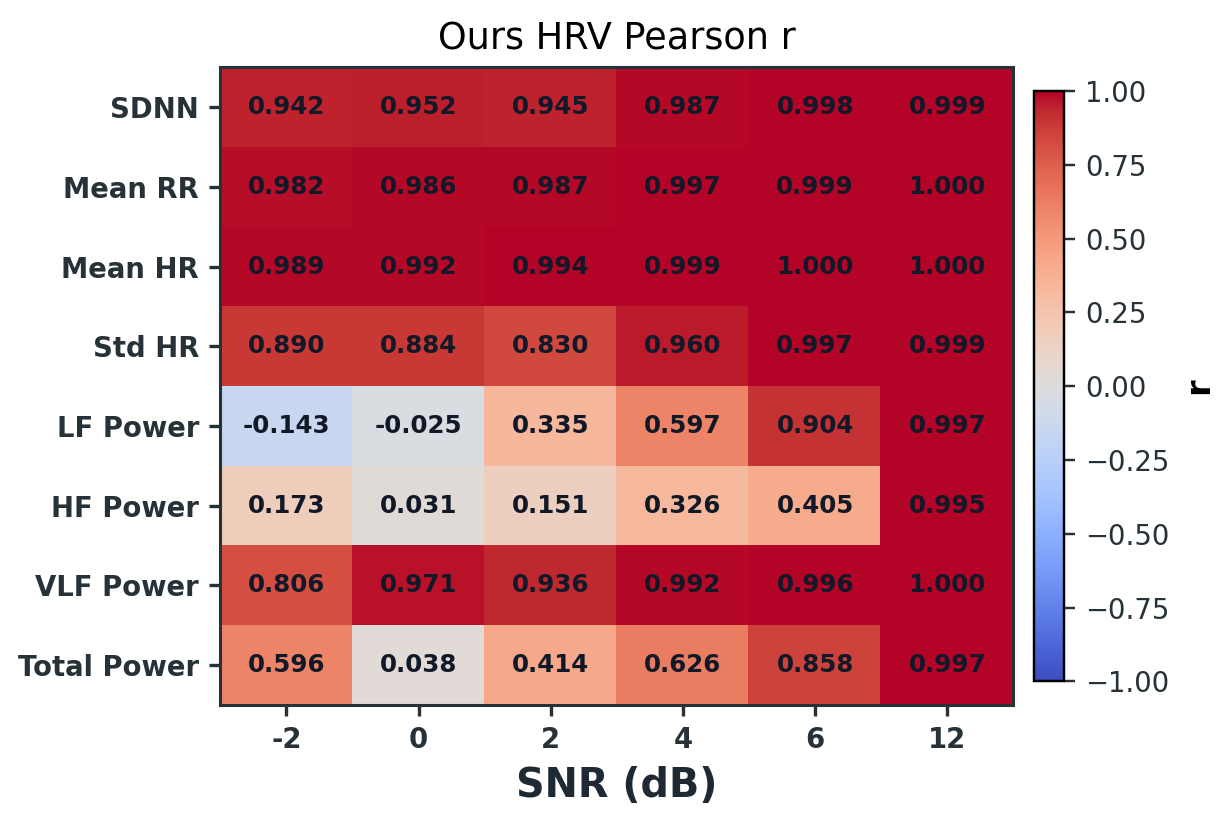}
    \vspace{-2mm}
    \caption{HRV metrics across SNR levels on SPC 2015.}
    \label{fig:hrv_snr_pearson}
    \vspace{-3mm}
\end{figure}

\subsection{Robustness Across Noise Levels and Subjects}
~\figref{snr_full} presents all four window-level metrics as a function of SNR level from -2 dB to 12 dB. ARTA remains stable as motion noise increases, while competing methods degrade more sharply at lower SNRs. Although WFDB-XQRS is marginally better at the cleanest SNR condition, ARTA provides the best overall robustness across the full noise range. 
At $12$~dB---where motion artifact is mild---ARTA and WFDB-XQRS perform nearly identically ($\mathrm{MAE}$ $0.519$ vs.\ $0.502$~BPM), confirming that the additional complexity of graph-based optimization is not needed when single-detector output is already reliable. The picture changes sharply as noise increases. Moving from $12$~dB to $0$~dB, XQRS MAE grows by a factor of $7.8\times$, Pan--Tompkins by $12.3\times$, and Wavelet by $15.5\times$, while ARTA grows by only $3.97\times$. The advantage of combining multi-detector candidates with rhythm-aware DAG optimization and a time-varying HR prior is most pronounced at 0 dB and -2 dB, where single-detector pipelines and prior-guided graph methods degrade most severely.
At $-2$~dB, the most challenging condition, ARTA achieves $\mathrm{MAE} = 3.89$~BPM compared to XQRS at $9.18$~BPM---a $57.6\%$ reduction.

~\figref{snr_boxplot_IBI} shows the distribution of IBI RMSE across SNR levels for all methods, where each box summarizes subject-level variability at a given noise condition. ARTA consistently achieves the lowest median IBI RMSE and one of the smallest interquartile ranges across all SNR conditions, indicating both high accuracy and low subject-to-subject variability. Although WFDB-XQRS improves at higher SNRs, ARTA remains more compact and stable over the full noise range, confirming that graph-based rhythm optimization reduces both median error and cross-subject variability.

\subsection{HRV Parameter Estimation}
~\tblref{hrv_pearson} and~\figref{hrv_snr_pearson} present Pearson correlations between ARTA-estimated and clean-ECG-derived values for eight HRV parameters (four time-domain and four frequency-domain) across different SNR levels. The results reveal a consistent two-tier structure across all methods: time-domain parameters (Mean RR, Mean HR, SDNN, STD HR) are substantially more robust to motion artifact than frequency-domain parameters (VLF, LF, HF, Total Power). Pearson correlations for Mean RR and Mean HR remained very high even at $-2$~dB ($r = 0.982$ and $r = 0.989$ respectively), and approached near-perfect agreement at $12$~dB ($r > 0.999$, $p < 0.001$). 
Frequency-domain HRV metrics are more sensitive to residual timing errors across all methods. LF, HF, and total power show weaker correlations at low SNRs, but their agreement improves as the signal becomes cleaner. This trend is expected because frequency-domain HRV depends on the spectral structure of the full IBI sequence and is therefore more vulnerable to occasional missed or false beats


\begin{figure}[h]
    \centering
    \includegraphics[width=1\linewidth]{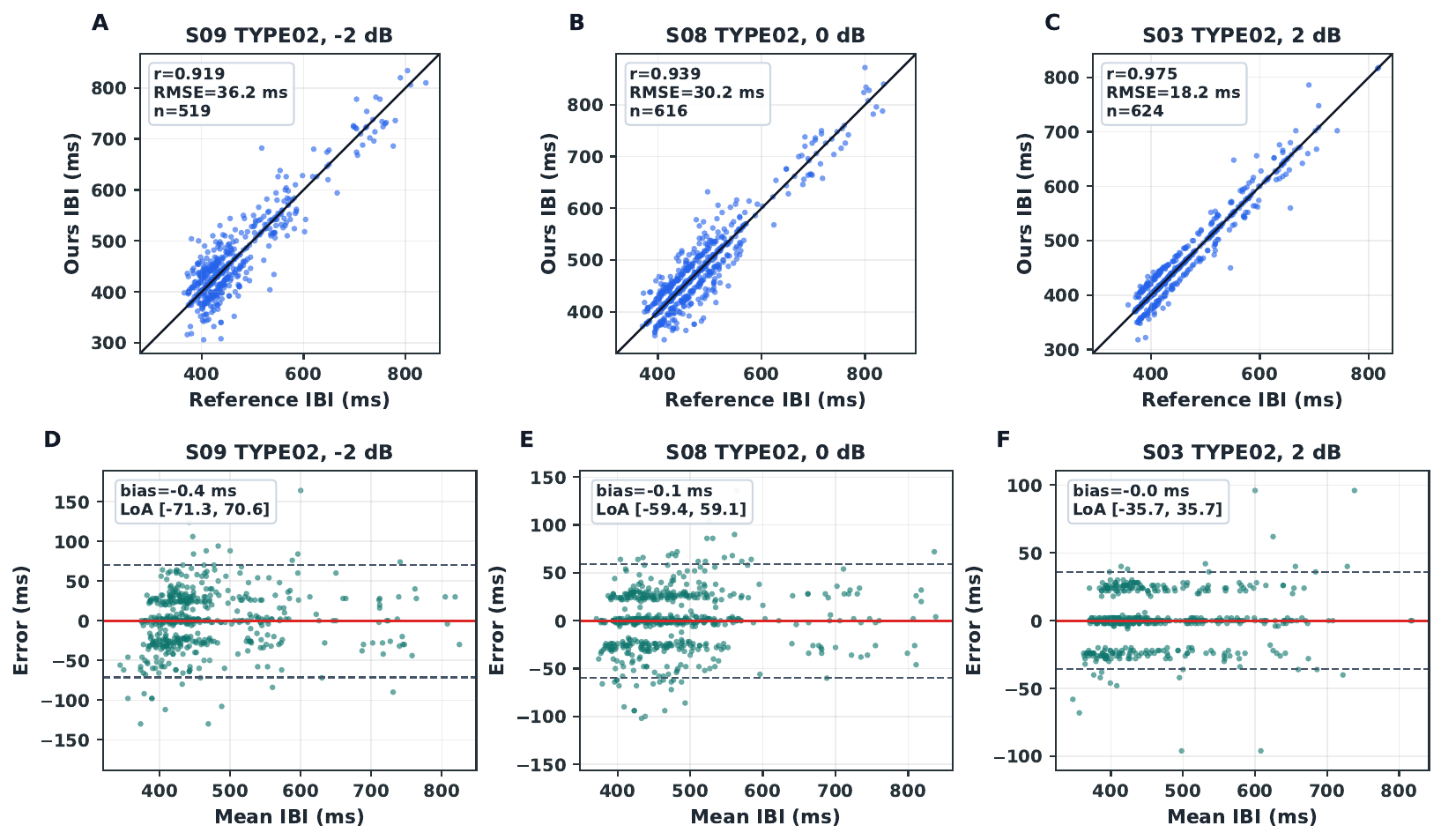}
    \caption{IBI agreement for representative SPC 2015 subject--SNR conditions: (a--c) Pearson agreement and (d--f) Bland--Altman analysis.}
    \label{fig:ibi_agreement}
\end{figure}
\subsection{Subject-Level IBI Agreement}
~\figref{ibi_agreement} shows Pearson scatter (a-c) and Bland-Altman (d-f) plots for three representative subject-conditions spanning -2 to 2 dB. The Pearson correlation plots show that estimated IBIs remain strongly correlated with reference IBIs even under severe motion artifact in all three cases. Even at -2 dB, the IBI estimates remain centered around a small bias, and agreement improves as SNR increases. The corresponding Bland-Altman plots show errors centered near zero bias, indicating that ARTA introduces no systematic timing shift. This is an important property for HRV applications, where a consistent offset between estimated and reference IBIs would bias SDNN and RMSSD calculations even if individual intervals are tightly distributed. The near-zero bias across all tested subjects and noise levels suggests the graph-based path optimization preserves the statistical distribution of IBI rather than merely minimizing peak-level error.

\begin{figure}
    \centering
    \includegraphics[width=1.02\linewidth, trim={6 8 8 35},clip]{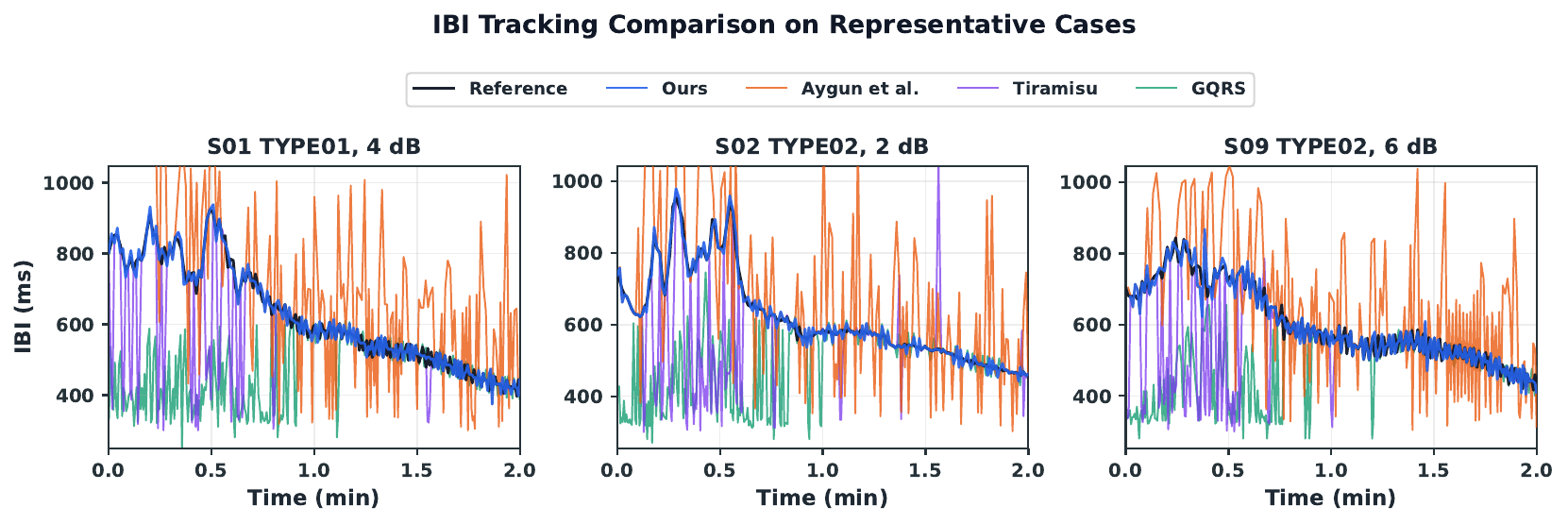}
    \caption{IBI Tracking comparison on representative cases.}
    \label{fig:ibi_time}
\end{figure}

In~\figref{ibi_time}, ARTA closely follows the temporal IBI trajectory, while competing methods show larger deviations, missed-beat jumps, and unstable intervals. This distinction matters clinically because a single 500 ms IBI error (a missed beat at 100 BPM) can corrupt an entire 8-second HRV window, whereas ten 20 ms errors distributed across the same window have negligible impact on SDNN or RMSSD. The graph-based optimization is not minimizing the number of individual errors but the global deviation of the path from a physiologically consistent rhythm, which is a more appropriate objective for HRV-quality IBI estimation.

\subsection{Ablation Study}
We isolate the contribution of each model component and characterize sensitivity to the two principal hyperparameters.~\tblref{ablation_loss} reports component ablations;~\tblref{sensitivity_beam_memory} and Figures~\ref{fig:beam_rhythm_memory}-~\ref{fig:hyper_parameter} report hyperparameter sensitivity.

\subsubsection{Hyperparameter Analysis of Cost Term}
\begin{table}[h]
\caption{Ablation study. Deltas shown relative to the full model.
         $n = 10{,}606$ windows except w/o $C_\text{ep}$+$C_\text{cov}$
         ($n=652$; path fragment collapse).}
\label{tab:ablation_loss}
\centering
\setlength{\tabcolsep}{2pt}
\begin{tabular}{cccccccc}
\toprule
$C_\text{node}$ & $C_\text{HR}$ & $C_\text{ep}$ & $C_\text{cov}$
        & MAE$\downarrow$ & RMSE$\downarrow$ & $r\uparrow$ & $W_{\pm5}(\%)\uparrow$ \\
\midrule
\checkmark & \checkmark & \checkmark & \checkmark
  & \textbf{1.534} & \textbf{4.618} & \textbf{0.982} & \textbf{93.61} \\
\midrule
$\times$ & \checkmark & \checkmark & \checkmark
  & 1.56\,{\scriptsize(+0.02)} & 4.66\,{\scriptsize(+0.04)} & 0.98\,{\scriptsize(0.00)} & 93.51\,{\scriptsize($-$0.10)} \\
\checkmark & $\times$ & \checkmark & \checkmark
  & 14.83\,{\scriptsize(+13.29)} & 34.32\,{\scriptsize(+29.70)} & 0.57\,{\scriptsize($-$0.42)} & 78.96\,{\scriptsize($-$14.65)} \\
\checkmark & \checkmark & $\times$ & \checkmark
  & 1.53\,{\scriptsize($-$0.002)} & 5.35\,{\scriptsize(+0.73)} & 0.97\,{\scriptsize($-$0.01)} & 93.96\,{\scriptsize(+0.35)} \\
\checkmark & \checkmark & \checkmark & $\times$
  & 1.53\,{\scriptsize(0.00)} & 4.62\,{\scriptsize(0.00)} & 0.98\,{\scriptsize(0.00)} & 93.61\,{\scriptsize(0.000)} \\
\checkmark & \checkmark & $\times$ & $\times$
  & 2.81\,{\scriptsize(+1.28)} & 12.76\,{\scriptsize(+8.15)} & 0.87\,{\scriptsize($-$0.11)} & 93.71\,{\scriptsize(+0.10)} \\
\bottomrule
\end{tabular}
\footnotesize{$^\dagger$$C_\text{local}$ retained in all variants; see~\secref{local_rhythm}}
\end{table}
~\tblref{ablation_loss} examines the contribution of each cost term by removing one component at a time while keeping all others at the full-model configuration. Removing the soft HR-prior term ($C_{\mathrm{HR}}$) causes the most severe degradation, increasing RMSE 643\%  and collapsing correlation from 0.982 to 0.57. This confirms that the time-varying HR prior is essential for guiding the graph search through ambiguous noisy candidates, particularly when multiple plausible paths exist at low SNR. Without this term, the optimizer has no global physiological anchor and can select rhythmically consistent but clinically implausible beat sequences. For example, a path running at 100 BPM throughout and a path running at 60 BPM throughout may both look locally consistent but only one is physiologically plausible for a given recording. The HR prior resolves this ambiguity by providing a soft global reference that guides the optimizer toward the correct operating range.

The endpoint and coverage penalties ($C_{\mathrm{ep}}$, $C_{\mathrm{cov}}$) play a structural rather than an accuracy role: removing both triggers path-fragment collapse and the optimizer selects a short, clean 5-second fragment of the recording, reducing evaluated windows from 10,606 to 652 (only 6.1\% coverage). This result show that without boundary constraints, the beat-reward term alone is insufficient to prevent the optimizer from exploiting the path-cost structure to find clean fragments rather than complete recordings.

Removing the candidate confidence term $C_{\mathrm{node}}$ produces only a small degradation ($\Delta\mathrm{MAE} = +0.022$~BPM, $\Delta W_{\pm 5} = -0.10$~pp), indicating that once candidates are generated, the graph-level rhythm constraints dominate the final selection. Candidate confidence is primarily a secondary tie-breaking term that refines the solution without controlling it.

\subsubsection{Beam width and rhythm memory}
\begin{figure}[h]
\centering
    \subfloat[RMSE versus SNR for $B = 2$ and $B = 3$.]{
    \includegraphics[width=0.51\linewidth, trim={2 3 3 10},clip]{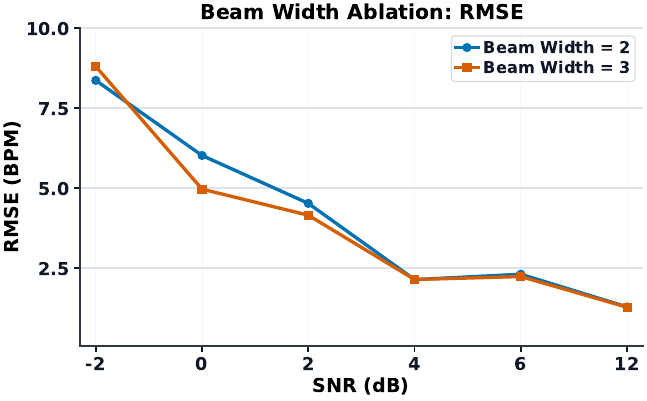}
    \label{fig:beam_size}
    }
    \subfloat[RMSE versus SNR for window sizes]{
    \includegraphics[width=0.51\linewidth, trim={2 4 4 13},clip]{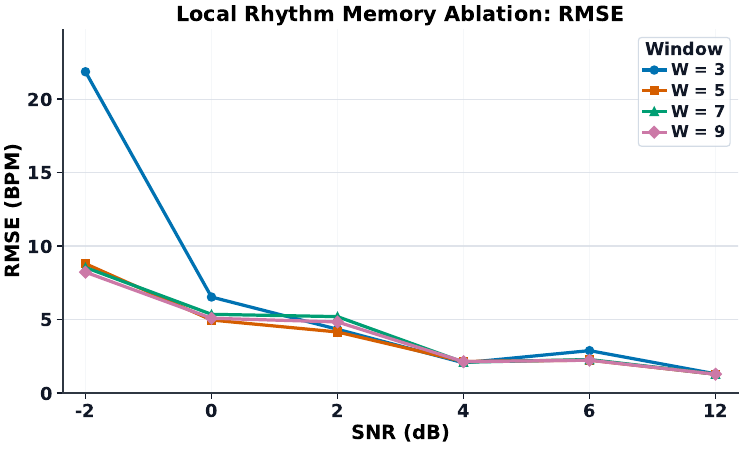}
    \label{fig:memory_size}
    }
    \caption{Sensitivity to beam width $B$ and local rhythm memory $\Omega$.(a) RMSE across SNR levels for beam widths $B = 2$ and $B = 3$. (b) RMSE across SNR levels for rhythm memory windows $\Omega = 3, 5, 7, 9$.}
     \label{fig:beam_rhythm_memory}
\end{figure}

~\tblref{sensitivity_beam_memory} and~\figref{beam_rhythm_memory} examine sensitivity to beam width and local rhythm memory size. Increasing the beam width from $B = 2$ to $B = 3$ gives a modest improvement, reducing 1.94\% RMSE from $5.315$ to $5.212$~BPM and slightly improving Pearson correlation and $W_{\pm 5}$. This suggests that the method does not require a large beam; a compact beam is sufficient because physiological RR constraints and rhythm costs already restrict the search space effectively. $B = 3$ was selected as the final configuration because it provides a consistent accuracy gain while keeping the dynamic programming search computationally manageable.
\begin{table}[]
\caption{Sensitivity analysis for beam width $B$ and local rhythm
         memory size $\Omega$.}
\label{tab:sensitivity_beam_memory}
\centering
\begin{tabular}{llcccc}
\toprule
& & MAE$\downarrow$ & RMSE$\downarrow$ & $r\uparrow$ & $W_{\pm5}(\%)\uparrow$ \\
\midrule
\multirow{2}{*}{\rotatebox[origin=c]{90}{\scalebox{0.85}{\shortstack{Beam\\size $B$}}}}
& $B=2$ & 1.822 & 5.315 & 0.976 & 92.24 \\
& $B=3$ & \textbf{1.796} & \textbf{5.212} & \textbf{0.977} & \textbf{92.38} \\
\midrule
\multirow{4}{*}{\rotatebox[origin=c]{90}{\scalebox{0.85}{\shortstack{ window\\size $\Omega$}}}}
& $\Omega=3$ & 3.189 & 11.519 & 0.889 & 90.16 \\
& \textbf{$\Omega=5$ }& \textbf{1.796} & \textbf{5.212} & \textbf{0.977} & \textbf{92.38} \\
& $\Omega=7$ & 1.795 & 5.206 & 0.977 & 92.07 \\
& $\Omega=9$ & 1.732 & 5.001 & 0.979 & 92.53 \\
\bottomrule
\end{tabular}
\end{table}

The local rhythm memory size has a stronger effect. A short memory ($\Omega=3$) performs noticeably worse, as the median and MAD are estimated from only two intervals, which means a single corrupted IBI can shift the median by 50\% and inflate the MAD arbitrarily. Increasing the memory size to 5 reduces the RMSE by ~50\%. Further increases to $\ Omega=7$ and $\Omega=9$ provide incremental gains (RMSE 5.206 and 5.001 BPM, respectively), with $\Omega=9$ achieving the best overall RMSE. The improvement beyond $\Omega=5$is relatively small and it provides a practical balance between responsiveness to genuine HR changes and robustness against noisy individual IBI estimates. 
\begin{figure*}[h]
\centering
    \subfloat[\textbf{Per-weight sensitivity}. RMSE as a function of each cost weight varied individually around the selected configuration (red star). The HR-prior and local-rhythm weights show the strongest effects.]{
    \includegraphics[width=0.81\textwidth, trim={6 8 6 5},clip]{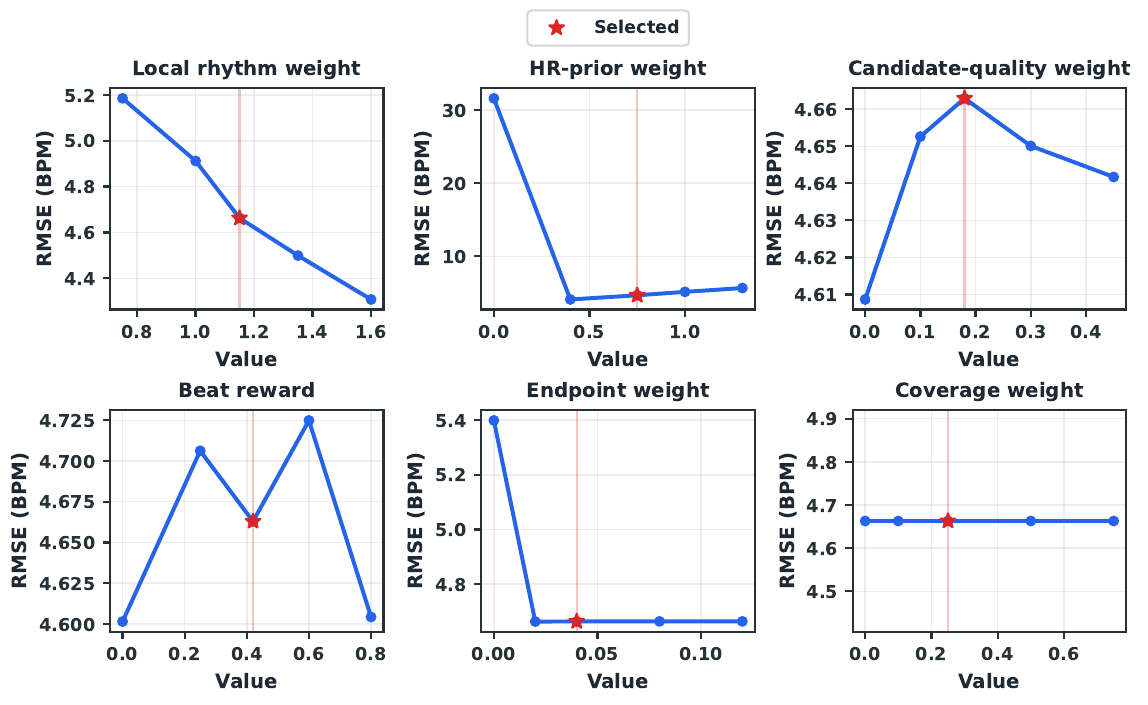}
    \label{fig:rmse_hyperparameter}
    }
    
    \subfloat[\textbf{RMSE heatmap.} Joint RMSE across all hyperparameter values (columns) and SNR levels (rows). The selected configuration lies within a broad low-error region rather than at an isolated optimum]{
    \includegraphics[width=0.9\textwidth, trim={4 8 6 18},clip]{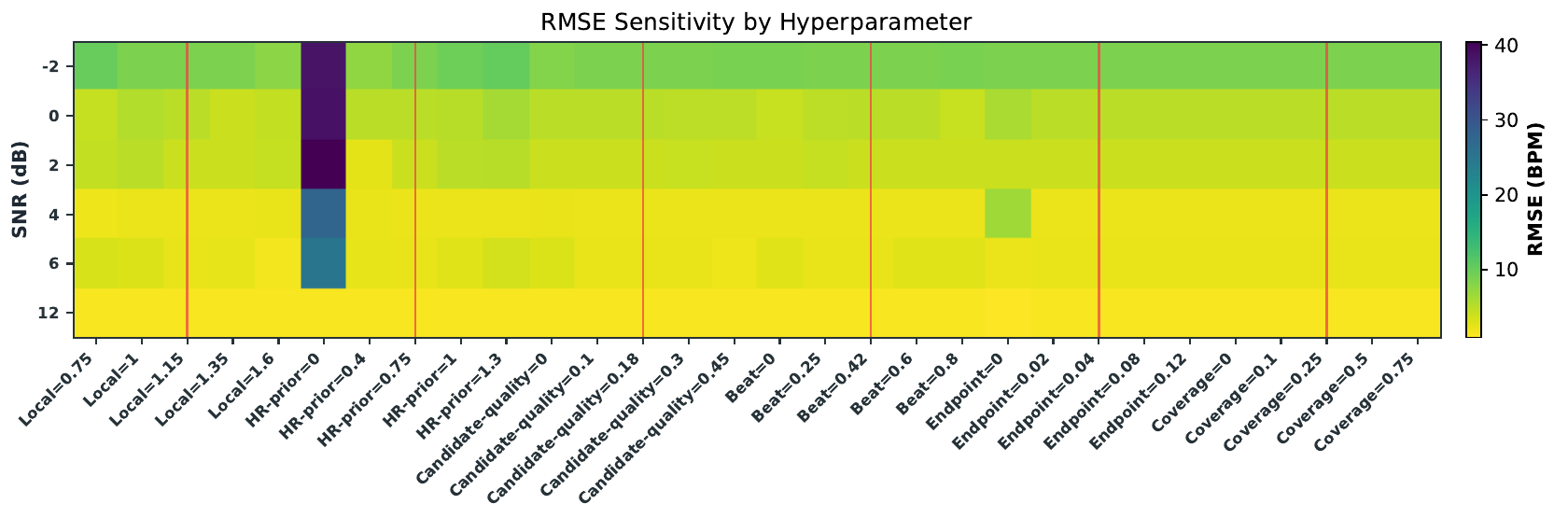}
    \label{fig:snr_heatmap_hyperparameter}
    }
     \caption{Hyperparameter sensitivity analysis. (a) RMSE as each cost weight is swept around its selected value (red star). (b) Joint RMSE landscape across hyperparameters and SNR levels.}
     \label{fig:hyper_parameter}
\end{figure*}

Overall, the ablation confirms that the HR prior and local rhythm memory are the most important components for robust graph-based IBI estimation under motion artifact.

\subsubsection{Robustness of the weight configuration}
~\figref{hyper_parameter} presents the sensitivity of overall RMSE to each cost weight individually (~\figref{rmse_hyperparameter}) and as a joint SNR-stratified heatmap (~\figref{snr_heatmap_hyperparameter}). The selected configuration is marked by a red star in each panel. Together they confirm that the cost function is not an ad hoc collection of tuning parameters but a structured set of components with distinct physiological roles and noise-level dependencies.

The HR-prior weight dominates the sensitivity: driving it to zero spikes RMSE, especially at low SNR, echoing the Table~\ref{tab:ablation_loss} ablation---confirming that the time-varying HR prior is essential for preventing the graph search from selecting false beat paths under strong motion artifact. The heatmap (~\figref{snr_heatmap_hyperparameter}) shows this effect is most pronounced at $-2$ and $0$~dB SNR, where the prior provides the most critical guidance when local rhythm history alone is insufficient to distinguish true beats from artifact.

The local-rhythm weight is the next most influential, with larger values lowering RMSE. The other weights (candidate-quality, beat-reward, endpoint, and coverage) have only marginal effect on RMSE, with total variation in RMSE below 0.15 BPM, consistent with their regularizing role. This confirms the hierarchy established by the cost-term ablation: these are structural regularization terms that prevent degenerate solutions, not precision-tuning parameters. The practical benefit is that the method can be deployed with approximately correct values for these parameters without careful calibration, provided $C_{\mathrm{prior}}$ and $C_{\mathrm{rhythm}}$ are properly weighted.

\begin{figure}
    \centering
    \includegraphics[width=1\linewidth, trim={0 0 0 0},clip]{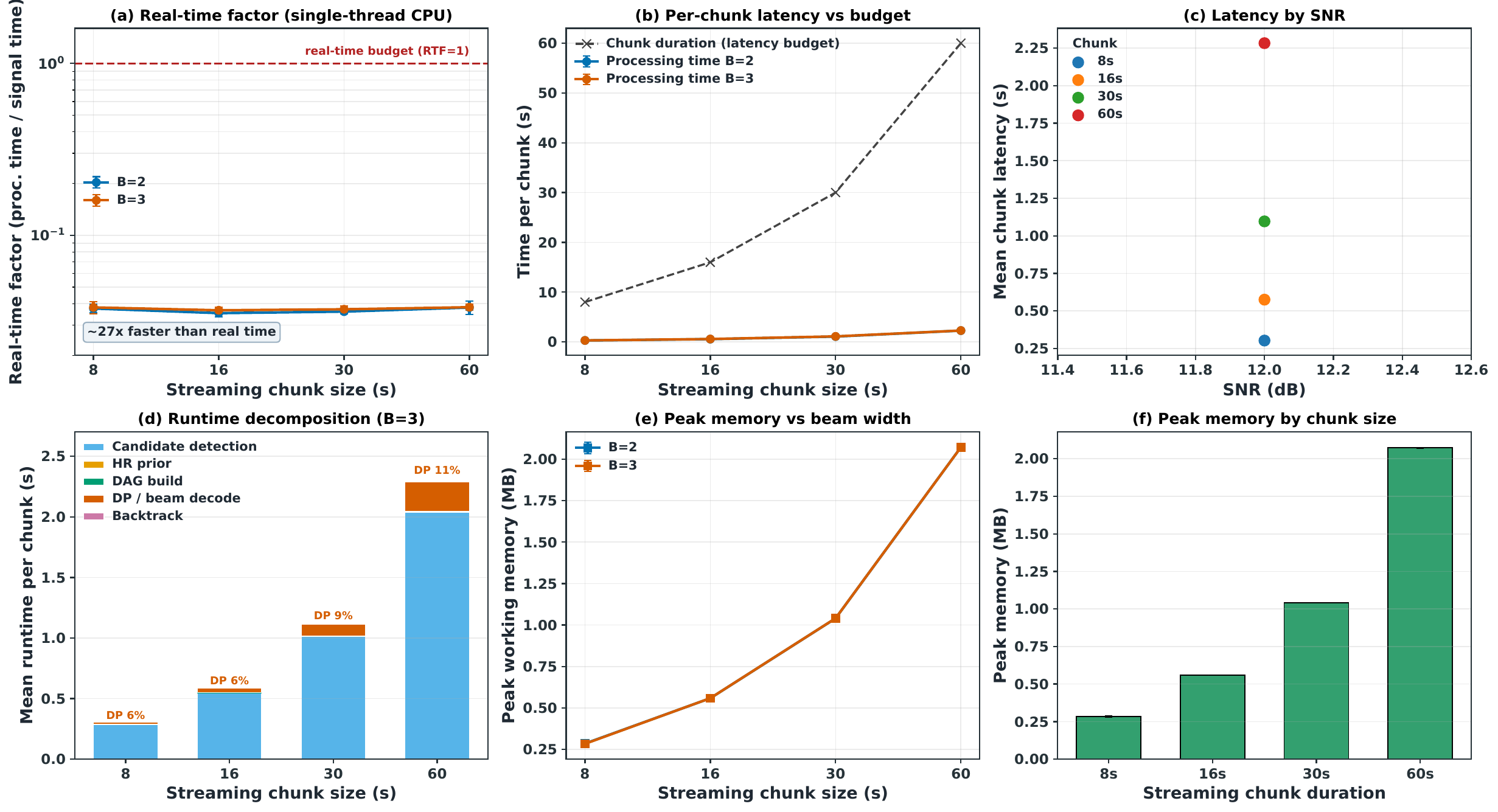}
    \caption{Single-thread CPU edge simulation. (a) Real-time factor across chunk sizes
  for beam widths $B{=}2$ and $B{=}3$. (b) Processing latency vs.\ available chunk
  budget. (c) Latency across simulated SNR levels. (d) Runtime decomposition by
  algorithmic stage. (e--f) Peak memory usage vs.\ beam width and chunk duration.}
  \label{fig:edge-sim}
\end{figure}

\subsubsection{Edge-Device Feasibility}
\label{sec:edge-feasibility}
To evaluate whether ARTA is compatible with resource-limited deployment, we performed a
single-thread CPU edge simulation, in which numerical parallelism was disabled and the
ECG signal was processed in streaming chunks of 8, 16, 30, and 60~s. This experiment
should be interpreted as a constrained CPU emulation rather than a direct smartwatch
benchmark. Results are summarized in Fig.~\ref{fig:edge-sim}(a--f).

Fig.~\ref{fig:edge-sim}(a) shows that ARTA remains far below the real-time limit for all
chunk sizes. The average real-time factor was approximately $0.037$, meaning the method
processed ECG roughly $27\times$ faster than real time. Increasing the beam width from
$B{=}2$ to $B{=}3$ produced only a small runtime increase, about $0.5$--$4\%$ depending
on chunk length, indicating that the final beam setting adds negligible streaming
overhead.

Fig.~\ref{fig:edge-sim}(b) compares processing time with the duration of each ECG chunk. For real-time operation, processing must finish before the next chunk arrives. For example, an 8 s chunk must be processed in less than 8 s. ARTA required only about 0.30 s for an 8 s chunk, using less than 4\% of the available time. For 60 s chunks, processing time increased to about 2.3 s, but this was still much smaller than the 60 s available duration. This indicates that runtime increases with chunk length but remains well below the requirement for real-time processing. Fig.~\ref{fig:edge-sim}(c) further shows that this real-time margin is maintained across SNR levels. Runtime is driven mainly by chunk duration rather than noise severity: longer ECG chunks contain more candidate peaks and graph edges, while different SNR levels produce only modest runtime variation.


Fig.~\ref{fig:edge-sim}(d) decomposes runtime by algorithmic stage. Candidate generation
dominates total runtime, while DAG construction, HR-prior estimation, backtracking, and
beam-search decoding contribute much smaller fractions. This suggests that the
graph-constrained decoding component itself is not the computational bottleneck.

Fig.~\ref{fig:edge-sim}(e--f) shows memory behavior. Peak memory remains low and is
nearly unchanged when increasing beam width from $B{=}2$ to $B{=}3$. Memory increases
with chunk duration but stays within only a few megabytes, rising from roughly
$0.29$~MB for 8~s chunks to about $2.07$~MB for 60~s chunks.

Overall, the edge simulation shows that ARTA operates substantially faster than real
time in a constrained single-thread CPU setting, with low memory demand and minimal
additional cost from the deployed beam width. This supports the feasibility of ARTA as
a lightweight, training-free heartbeat tracking method for wearable-style ECG
processing.

\section{Discussion, Limitation and Future Work}




The results show that ARTA improves ECG-based HR and IBI estimation under motion artifact by treating heartbeat detection as a global sequence-decoding problem rather than a set of independent peak-level decisions. This is important because motion artifact creates two coupled errors: false QRS-like peaks and missed true beats. By constructing an over-complete candidate set from weak detectors and selecting the most physiologically consistent path through the graph, ARTA can reject false positives and recover missed beats within the same optimization framework. The balanced improvement in precision and recall indicates that the method is not simply changing detection sensitivity, but is improving the quality of the selected heartbeat sequence.

The ablation and sensitivity results further suggest that both global and local constraints are necessary. The soft HR prior helps guide the graph when the signal is highly ambiguous, while the path-local rhythm memory allows the method to adapt to the recent IBI pattern rather than relying only on a global average-HR estimate. Endpoint and coverage penalties are also important because they prevent the optimizer from selecting only a short clean fragment of the recording. Together, these components allow ARTA to preserve beat-to-beat timing accuracy and produce reliable window-level HR estimates, while remaining interpretable and training-free compared with deep-learning approaches.

The EDB evaluation highlights the difference between robustness to controlled additive artifact and generalization to clinical ambulatory ECG. Although ARTA maintained a window-level advantage over the evaluated comparators, its lower absolute accuracy shows that abnormal morphology, ectopy, heterogeneous leads, nonstationary rhythms, and native signal-quality changes remain challenging. Importantly, ARTA's label-free diagnostics helped distinguish more reliable segments without using reference annotations, suggesting that low-confidence outputs could be flagged or routed to alternative processing. However, this analysis was post hoc and its normalization bounds were estimated from EDB. Independent prospective calibration is therefore required before the score can be used as an operational rejection rule.

This study has several limitations. First, beat-level evaluation on IEEE SPC used clean-ECG-derived reference peaks rather than manual expert annotations, so the results should be interpreted as agreement with a clean reference rather than clinical ground truth, whereas EDB evaluation used expert-reviewed beat annotations. Second, ARTA can only recover beats that appear in the candidate set; if all weak detectors miss a true R-peak, the graph cannot reconstruct it. The over-complete detector design reduces this risk but does not eliminate it. Third, although EDB extends the evaluation to long clinical ambulatory ECG, its development subset was used to select the rhythm-memory length, and the label-free reliability analysis was post hoc. Independent evaluation on an untouched clinical dataset is therefore required to establish prospective generalization and calibrate a fixed reliability threshold. Future work will focus on signal-quality-aware candidate generation, arrhythmia-aware graph constraints, and real-time deployment on wearable platforms.

\section{Conclusion}
This work presented ARTA (Artifact-Robust Temporal Alignment), a graph  combinatorial optimization framework for robust ECG-based IBI and heart-rate estimation under motion artifact. The key novelty is to treat outputs from multiple ECG detectors as weak candidate evidence and recover the final R-peak sequence through global rhythm-aware optimization rather than independent peak selection, heuristic repair, or majority voting. By constructing an over-complete candidate DAG and selecting the minimum-cost path using beam-search dynamic programming, ARTA combines evidence from candidate peaks with physiological rhythm constraints to identify the most plausible heartbeat sequence across the full recording. This formulation enables false-positive rejection and missed-beat recovery within a single interpretable optimization framework. Importantly, ARTA is training-free and computationally lightweight, requiring no labeled data or externally sensed signals — properties that make it directly deployable on resource-constrained wearable platforms. By delivering accurate beat-to-beat IBI and heart-rate estimates that remain reliable under severe motion, ARTA provides a robust foundation for the broad range of downstream applications that depend on precise cardiac timing, including continuous cardiovascular risk monitoring, HRV-based stress and sleep assessment, fitness and exercise tracking, and affective computing — extending reliable cardiac monitoring from the clinic into everyday free-living conditions.


\section*{Data \& Code Availability}
The codebase is publicly available at: \href{https://github.com/shovito66/arta}{\textcolor{blue}{https://github.com/shovito66/arta}}.
The interactive dashboard for testing the proposed method is available at: \href{https://arta.ghasemzadeh.com/}{\textcolor{blue}{https://arta.ghasemzadeh.com}}


\begin{acks}
This work was supported in part by the National Science Foundation (NSF) under grant 2531465. The content is solely the responsibility of the authors and does not necessarily represent the official views of the NSF.
\end{acks}

\appendix
\section{Definition of the ARTA-Derived Reliability Score}
\label{app:arta_reliability}

The selective-reliability analysis used quantities generated internally by
ARTA from the noisy ECG. It did not use reference R-peaks, BPM labels,
beat-detection F1, BPM error, IBI error, or any other ground-truth quantity.
The four diagnostics were calculated separately for every 120-s record--SNR
segment instance. Throughout this appendix,
$\operatorname{clip}(x,0,1)=\min(1,\max(0,x))$.

\subsection{Mean Local Candidate SQI}

For each merged R-peak candidate, ARTA calculated a local signal-quality value
from a QRS-preserving ECG signal filtered between 0.5 and 40~Hz. Let
$p_{i,j}$ denote candidate $j$ in segment instance $i$. Local background
variation was measured within 250~ms of the candidate after excluding samples
within 90~ms of its center. A robust local noise scale was estimated using the
median absolute deviation (MAD):

\begin{equation}
\sigma_{i,j}
=
1.483\,
\operatorname{median}_{u\in\mathcal{N}_{i,j}}
\left|
u-\operatorname{median}(\mathcal{N}_{i,j})
\right|
+\epsilon,
\end{equation}

\noindent where $\mathcal{N}_{i,j}$ contains the local background samples
outside the excluded QRS-centered interval. Candidate prominence was then

\begin{equation}
r_{i,j}
=
\frac{
\left|x_{\mathrm{QRS}}[p_{i,j}]\right|
}{
\sigma_{i,j}
}.
\end{equation}

The prominence ratio was logarithmically compressed and bounded to $[0,1]$:

\begin{equation}
s_{i,j}^{\mathrm{SQI}}
=
\operatorname{clip}
\left(
\frac{
\log(1+r_{i,j})-\log(1+2)
}{
\log(1+20)-\log(1+2)
},
0,1
\right).
\end{equation}

For the $M_i$ merged candidates in segment instance $i$, mean candidate SQI
was defined as

\begin{equation}
q_i^{\mathrm{SQI}}
=
\frac{1}{M_i}
\sum_{j=1}^{M_i}
s_{i,j}^{\mathrm{SQI}}.
\end{equation}

Higher values indicate that candidate peaks are, on average, more prominent
relative to their local ECG background. This is an ARTA candidate-level
diagnostic and not a clinically calibrated whole-record ECG quality index.

\subsection{Detector-Consensus Fraction}

ARTA generates candidate peaks using XQRS, Pan--Tompkins, and Wavelet
detectors. Detections within an 80-ms tolerance were considered to support the
same merged candidate. Let $d_{i,j}\in\{1,2,3\}$ denote the number of detector
families supporting merged candidate $j$. Detector-consensus fraction was

\begin{equation}
q_i^{\mathrm{cons}}
=
\frac{1}{M_i}
\sum_{j=1}^{M_i}
\mathbb{I}\left(d_{i,j}\geq 2\right),
\end{equation}

\noindent where $\mathbb{I}(\cdot)$ is the indicator function. The resulting
value is the fraction of merged candidates supported by at least two detector
families. Higher values indicate stronger agreement among the constituent QRS
detectors. Detector agreement should not, however, be interpreted as
statistically independent confirmation because the three detectors process the
same ECG signal.

\subsection{Mean HR-Prior Confidence}

ARTA derives a local HR prior from a robust subset of ECG candidates. Under
the robust-prior configuration, consensus-supported and XQRS candidate sets
were considered, and the sparser available set with at least two candidates
was used. If neither set was available, the merged candidate set was used as a
fallback.

For each local 8-s prior window $t$, four label-free properties were evaluated.
Let $n_{i,t}^{\mathrm{all}}$ be the number of merged candidates and
$n_{i,t}^{\mathrm{prior}}$ the number of prior candidates in the window.
Candidate-density consistency was defined as

\begin{equation}
D_{i,t}
=
\operatorname{clip}
\left(
\frac{
2.2-
n_{i,t}^{\mathrm{all}}/
\max(1,n_{i,t}^{\mathrm{prior}})
}{
1.2
},
0,1
\right).
\end{equation}

The local consensus fraction was

\begin{equation}
A_{i,t}
=
\frac{
\sum_{j\in\mathcal{C}_{i,t}}
\mathbb{I}(d_{i,j}\geq 2)
}{
|\mathcal{C}_{i,t}|
},
\end{equation}

\noindent where $\mathcal{C}_{i,t}$ is the set of merged candidates in the
local window.

Rhythm stability was calculated from physiologically valid prior-candidate RR
intervals between 300 and 2200~ms. If at least two such intervals were
available,

\begin{equation}
R_{i,t}
=
\operatorname{clip}
\left(
1-
\frac{
\operatorname{MAD}(\mathrm{RR}_{i,t})
}{
\max\left(
80~\mathrm{ms},
0.25\,\operatorname{median}(\mathrm{RR}_{i,t})
\right)
},
0,1
\right).
\end{equation}

When fewer than two valid RR intervals were available, $R_{i,t}$ was set to
0.45. Prior-candidate sufficiency was

\begin{equation}
U_{i,t}
=
\operatorname{clip}
\left(
\frac{n_{i,t}^{\mathrm{prior}}-1}{4},
0,1
\right).
\end{equation}

The local prior confidence was then

\begin{equation}
c_{i,t}
=
c_{\min}
+
(1-c_{\min})
\left(
0.40D_{i,t}
+
0.35A_{i,t}
+
0.25R_{i,t}
\right)
U_{i,t},
\end{equation}

\noindent with confidence floor $c_{\min}=0.25$. If no merged candidates or
fewer than two prior candidates were present, confidence was set directly to
$c_{\min}$. The local confidence sequence was smoothed using a three-window
moving average. Mean HR-prior confidence for segment instance $i$ was

\begin{equation}
q_i^{\mathrm{prior}}
=
\frac{1}{K_i}
\sum_{t=1}^{K_i}
c_{i,t},
\end{equation}

\noindent where $K_i$ is the number of local prior windows. Higher values
indicate that the ECG-derived rhythm prior was better supported by the
candidate data. This value is an internal ARTA confidence weight and not a
calibrated probability that the selected path is correct.

\subsection{Selected-Path Temporal Span}

Let $p_i^{(1)}$ and $p_i^{(L_i)}$ denote the first and last R-peaks in the
selected ARTA path, and let $N_i$ denote the number of samples in the 120-s
segment. The selected-path temporal span was

\begin{equation}
q_i^{\mathrm{span}}
=
100\,
\operatorname{clip}
\left(
\frac{
p_i^{(L_i)}-p_i^{(1)}
}{
N_i
},
0,1
\right).
\end{equation}

The value was set to zero when fewer than two peaks were selected. Higher values indicate that the selected trajectory extends from near the beginning to near the end of the segment. This quantity measures endpoint span rather than continuous path coverage; it does not explicitly identify large internal gaps between selected beats. 

\subsection{Robust Normalization and Ranking}

The four diagnostics have different numerical ranges. For diagnostic $k$, the
pooled 1st and 99th percentiles over all 32,400 EDB record--SNR segment
instances were denoted by $P_{1,k}$ and $P_{99,k}$. Each diagnostic was
normalized as

\begin{equation}
z_{i,k}
=
\operatorname{clip}
\left(
\frac{
q_{i,k}-P_{1,k}
}{
P_{99,k}-P_{1,k}
},
0,1
\right).
\end{equation}

The percentile values used in the analysis are reported in
Table~\ref{tab:quality_normalization_bounds}.

\begin{table}[t]
\centering
\caption{Percentile bounds used to normalize the EDB reliability diagnostics.}
\label{tab:quality_normalization_bounds}
\begin{tabular}{lcc}
\hline
Diagnostic & 1st percentile & 99th percentile \\
\hline
Mean candidate SQI              & 0.12924 & 0.73246 \\
Detector-consensus fraction     & 0.49157 & 0.99528 \\
Mean HR-prior confidence        & 0.46261 & 0.98913 \\
Selected-path temporal span (\%)& 58.73498 & 99.84500 \\
\hline
\end{tabular}
\end{table}

The final reliability score was the equally weighted average

\begin{equation}
Q_i
=
\frac{1}{4}
\left(
z_{i,\mathrm{SQI}}
+
z_{i,\mathrm{consensus}}
+
z_{i,\mathrm{prior}}
+
z_{i,\mathrm{span}}
\right).
\end{equation}

All record--SNR segment instances were ranked globally in descending order of $Q_i$. Metrics were recomputed after retaining the highest-scoring 100\%, 90\%, 80\%, 70\%, 60\%, 50\%, 40\%, and 30\% of instances. The corresponding retained counts were 32,400, 29,160, 25,920, 22,680, 19,440, 16,200, 12,960, and 9,720, respectively. Because global ranking was used, the retained proportion within an individual SNR condition was not necessarily equal to the nominal pooled retention level. The analysis therefore represents an exploratory performance--coverage sensitivity analysis rather than a prospectively calibrated segment-rejection rule.

\section{Supplementary EDB Rhythm-Memory Sensitivity Analysis}
\label{app:edb_memory}
Table~\ref{tab:edb_memory_sensitivity} evaluates the local rhythm-memory length on a fixed EDB development subset, with beam width $B=3$ and all other settings held constant. Compared with $\Omega=5$, using $\Omega=3$ reduced BPM MAE and RMSE by 14.8\% and 14.2\%, respectively, and increased Pearson correlation from 0.417 to 0.466. It also reduced IBI MAE by 2.2\%. Although longer memories provided a small improvement in $W_{\pm5}$, they increased the magnitude of the larger BPM errors. This behavior is consistent with a shorter memory adapting more rapidly to local rhythm changes and limiting the persistence of an incorrect trajectory after false candidates enter the graph. Therefore, $\Omega=3$ was selected for the EDB evaluation based primarily on MAE, RMSE, correlation, and IBI accuracy.
\begin{table}[h]
\centering
\caption{Sensitivity of ARTA to local rhythm-memory length on the
fixed EDB development subset. BPM and IBI errors are reported in BPM
and milliseconds, respectively.}
\label{tab:edb_memory_sensitivity}
\begin{tabular}{lrrrrr}
\toprule
Memory &
MAE$\downarrow$ &
RMSE$\downarrow$ &
$r\uparrow$ &
$W_{\pm5}$ (\%)$\uparrow$ &
IBI MAE$\downarrow$ \\
\midrule
$\Omega=3$ & \textbf{12.055} & \textbf{20.952} &
\textbf{0.466} & 58.16 & \textbf{23.480} \\
$\Omega=5$ & 14.151 & 24.422 & 0.417 & 58.23 & 23.999 \\
$\Omega=7$ & 14.723 & 25.614 & 0.375 & 59.00 & 24.190 \\
$\Omega=9$ & 14.473 & 25.603 & 0.380 &
\textbf{59.93} & 24.207 \\
\bottomrule
\end{tabular}
\end{table}

This sensitivity analysis was conducted on a fixed EDB development subset. Consequently, its absolute values differ from the primary full-dataset results reported in Table~\ref{tab:overall_comparison}. It is intended to compare memory settings under identical conditions, not to report final EDB performance.

\section{Supplementary Selective-Reliability Results}
\label{app:arta_reliability_results}

Table~\ref{tab:edb_reliability_operating_points} reports all evaluated reliability operating points. Segment retention denotes the fraction of the 32,400 record--SNR segments retained after global reliability-score ranking. Overall valid-window coverage denotes the number of valid BPM estimates from the retained segments divided by the total number of possible BPM windows before reliability-based exclusion. It therefore reflects both deliberate segment abstention and occasional missing BPM estimates within retained segments.

\begin{table}[h]
\centering
\caption{Complete label-free selective-reliability results on EDB. The
100\% row is the primary full-segment result; lower-retention rows represent
conditional performance at reduced coverage.}
\label{tab:edb_reliability_operating_points}
\small
\setlength{\tabcolsep}{5pt}
\begin{tabular}{rrrrrrrr}
\toprule
Segments retained &
Overall valid-window &
BPM MAE &
BPM RMSE &
Pearson $r$ &
$W_{\pm5}$ &
R-peak F1 &
IBI MAE \\
(\%) & coverage (\%) & (BPM) & (BPM) & & (\%) & & (ms) \\
\midrule
100 & 98.93 & 10.084 & 18.956 & 0.539 & 62.28 & 0.679 & 18.368 \\
 90 & 89.61 &  9.346 & 18.235 & 0.569 & 64.97 & 0.702 & 17.161 \\
 80 & 79.75 &  8.930 & 17.816 & 0.587 & 66.50 & 0.716 & 15.913 \\
 70 & 69.81 &  8.614 & 17.558 & 0.596 & 68.00 & 0.731 & 14.461 \\
 60 & 59.86 &  8.099 & 17.136 & 0.615 & 70.47 & 0.753 & 12.987 \\
 50 & 49.89 &  7.373 & 16.390 & 0.651 & 73.48 & 0.779 & 11.486 \\
 40 & 39.92 &  6.324 & 15.128 & 0.707 & 77.39 & 0.814 &  9.905 \\
 30 & 29.95 &  4.877 & 13.092 & 0.782 & 82.29 & 0.856 &  8.214 \\
\bottomrule
\end{tabular}
\end{table}

\begin{figure}[h]
    \centering
    \includegraphics[width=0.88\textwidth]
    {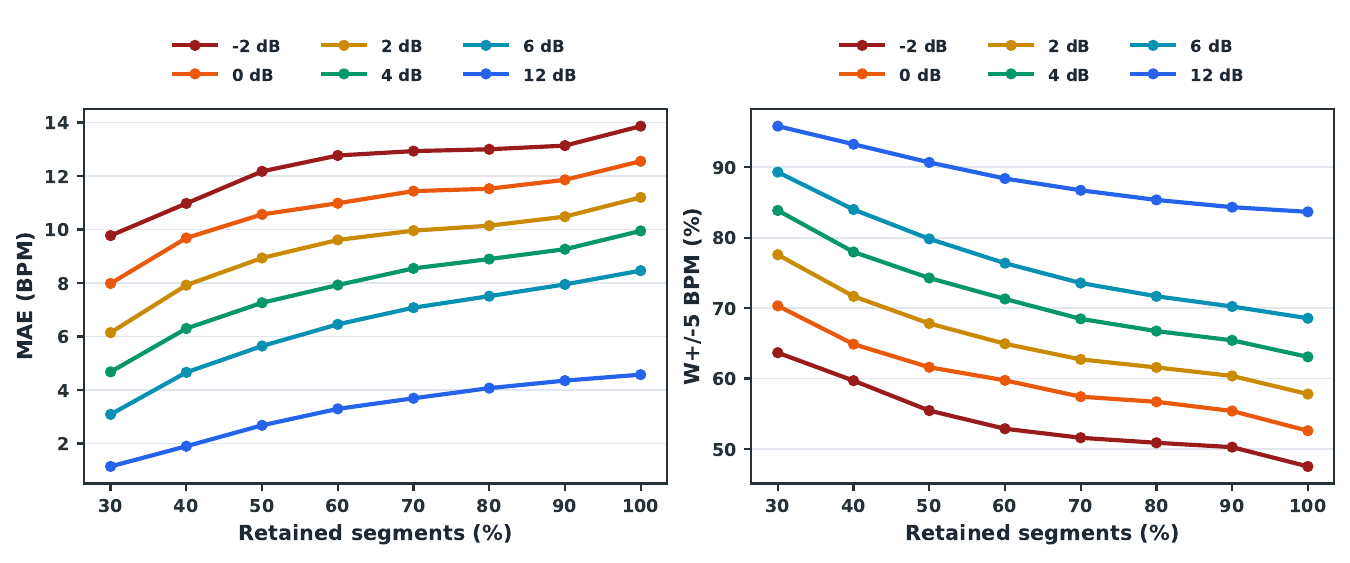}
    \caption{SNR-stratified label-free reliability sensitivity on EDB. BPM MAE and $W_{\pm5}$ are shown after retaining progressively larger proportions of the highest-scoring segment--SNR instances.}
    \label{fig:edb_reliability_by_snr}
\end{figure}

\begin{figure}[h]
    \centering
    \includegraphics[width=0.88\textwidth]
    {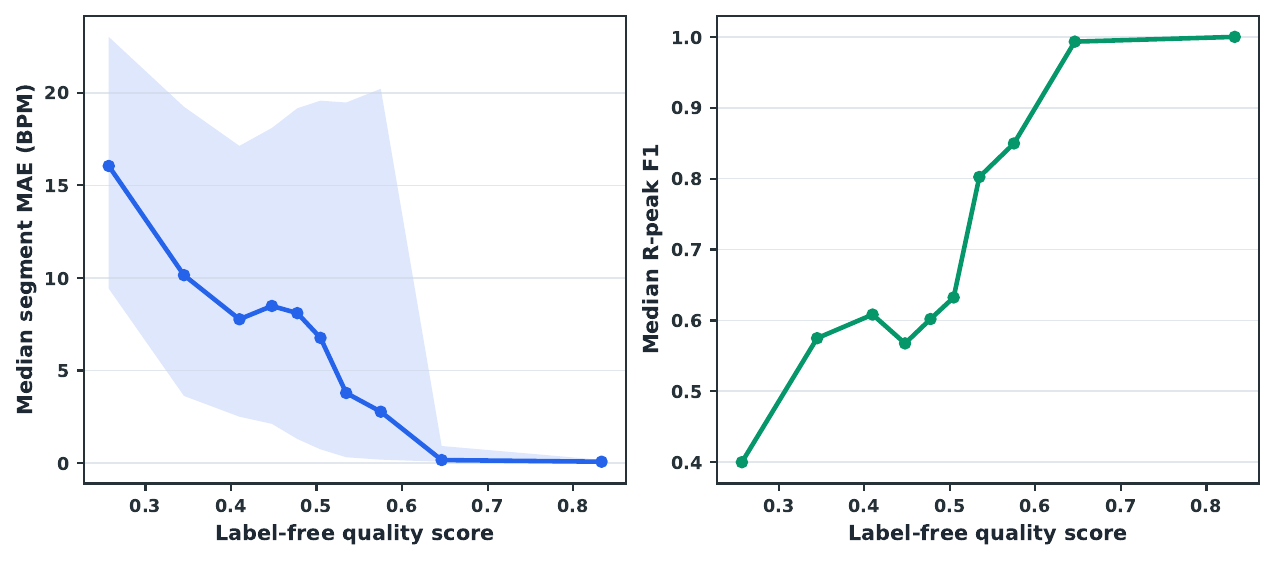}
    \caption{Relationship between the label-free ARTA reliability score and observed EDB accuracy. The left panel reports median segment-level BPM MAE, with the shaded region representing the interquartile range. The right panel reports median R-peak F1. Reference annotations were used only to evaluate this relationship and were not used to calculate the reliability score.}
    \label{fig:quality_score_error}
\end{figure}
Figure~\ref{fig:quality_score_error} examines the relationship between ARTA's label-free reliability score and observed accuracy. Segment--SNR instances were divided into reliability-score groups, and error metrics were calculated independently within each group. Higher reliability scores were generally associated with lower median BPM error and higher R-peak F1. The relationship was strongest in the highest-scoring groups, where candidate quality, detector agreement, HR-prior confidence, and path span were simultaneously favorable. The variability within intermediate groups indicates that the score provides a useful reliability indicator but is not a calibrated probability of correct estimation.

\bibliographystyle{ACM-Reference-Format}
\bibliography{ref}

\end{document}